\documentclass[preprint,12pt]{elsarticle}

\usepackage{amssymb}
\usepackage[margin=2.5cm]{geometry} 
\usepackage{lipsum}
\usepackage{graphicx}
\usepackage{caption}
\usepackage{subcaption}
\usepackage{diagbox}
\usepackage{multirow}
\usepackage{booktabs}
\usepackage{diagbox}
\usepackage{amsmath}

\usepackage[usenames,dvipsnames]{color}
\usepackage{multirow}
\usepackage{amssymb}
\usepackage{amsmath}
\usepackage{graphicx}
\usepackage{subcaption}
\usepackage{float} 
\usepackage{natbib}

\setcitestyle{square}

\begin{document}
\begin{frontmatter}



\title{Neural Correlates of Augmented Reality Safety Warnings: EEG Analysis of Situational Awareness and Cognitive Performance in Roadway Work Zones}


\author[inst1]{Fatemeh Banani Ardecani}
\author[inst1]{Amit Kumar}
\author[inst1]{Sepehr Sabeti}
\author[inst1]{Omidreza Shoghli\corref{cor1}}
\affiliation[inst1]{organization={William State Lee College of Engineering, University of North Carolina at Charlotte},
            addressline={9201 University City Blvd}, 
            city={Charlotte},
            postcode={28223}, 
            state={North Carolina},
            country={USA}}

\cortext[cor1]{Corresponding author: oshoghli@charlotte.edu}

\begin{abstract}

Despite the research and implementation efforts involving various safety strategies, protocols, and technologies, work zone crashes and fatalities continue to occur at an alarming rate each year. This study investigates the neurophysiological responses to Augmented Reality safety warnings in roadway work zones under varying workload conditions. Using electroencephalogram (EEG) technology, we objectively assessed situational awareness, attention, and cognitive load in simulated low-intensity (LA) and moderate-intensity (MA) work activities. The research analyzed key EEG indicators including beta, gamma, alpha, and theta waves, as well as various combined wave ratios. Results revealed that AR warnings effectively triggered neurological responses associated with increased situational awareness and attention across both workload conditions. However, significant differences were observed in the timing and intensity of these responses. In the LA condition, peak responses occurred earlier (within 125 ms post-warning) and were more pronounced, suggesting a more robust cognitive response when physical demands were lower. Conversely, the MA condition showed delayed peak responses (125-250 ms post-warning) and more gradual changes, indicating a potential impact of increased physical activity on cognitive processing speed.
These findings underscore the importance of considering physical workload when designing AR-based safety systems for roadway work zones. The research contributes to the understanding of how AR can enhance worker safety and provides insights for developing more effective, context-aware safety interventions in high-risk work environments.

\end{abstract}





\end{frontmatter}



\section{Introduction}
\label{sec1}

Roadway construction and maintenance are essential for sustaining highway performance, supporting national economic growth, and meeting the growing transportation demands of communities. Consequently, work zones are an integral aspect of roadway operation, maintenance, and construction activities, and play a fundamental role in advancing and supporting these socio-economic and transportation goals. However, despite the implementation and investigation of various safety strategies, protocols, and technologies, work zone crashes and fatalities continue to occur at an alarming rate each year. The Bureau of Transportation Statistics reports a consistent increase in work zone fatalities since 2010, with 2,066 fatal work injuries occurring in the transportation sector in 2022 alone \citep{BLS_CFOI_2023}.

To address safety issues in high-risk workplace environments, researchers have explored the use of innovative technologies such as Augmented Reality (AR) to manage safety hazards in these environments \citep{tatic2017application, gualtieri2023human, bavelos2024augmented}. 
In roadway work zones, these technologies are often intended to notify workers of potential intrusions and offer them additional time and information to respond to hazards \citep{sabeti2024augmented, sabeti2024mixed}. For this purpose, AR has particularly attracted researchers' interest because it can deliver real-time information overlays directly in workers' field of view, and enhance their understanding of safety risks \citep{
sitompul2019using, sitompul2020presenting}.

While the use of AR to improve workplace safety is expanding across various domains, there remains a significant gap in research that objectively assesses and quantifies the impact of AR-based safety systems on situational awareness, attention, and cognitive load. 
Situational awareness encompasses the perception of the elements in the environment over time and space, comprehending their meaning, and anticipating their near-future status \citep{endsley1995measurement}. 
With this, real-time safety systems, including AR-based solutions, are designed to enhance workers' real-time decision-making capabilities through increased situational awareness \citep{ibrahim2023investigating}, improve reaction times, and reduce information overload. 
This is particularly important in workplace safety, where research has shown that most accidents are caused by poor situational awareness \citep{ibrahim2023investigating}. In previous research, Caperllera et al. \citep{capallera2022human} analyzed the human-vehicle interaction to support driver's situational awareness in automated vehicles. Aromaa et al. \citep{aromaa2020awareness} assessed the industrial operators' awareness of the real-world environment when using augmented reality and Yu et al. \citep{yu2023air} determined the cognitive ergonomics-based AR application for construction performance. While these studies are valuable for advancing precedent understanding, some of these works lack objectivity and rely solely on subjective assessments, while others are not specific to roadway work zones. 

Furthermore, the effect of workload on situational awareness remains relatively under-explored in the context of roadway work zones. Physical workload has been shown to influence situational awareness \citep{kim2022construction,
zhang2023digital}. This is particularly relevant in roadway work zones, where workers are often engaged in physical tasks. The effects of high physical workload can lead to increased error rates, exhibiting unsafe behaviors, and diminished situational awareness \citep{chenarboo2022influence}. Such reductions in situational awareness can impair workers' ability to identify and analyze hazards, make projections, and maintain control, ultimately resulting in poor safety decisions \citep{ibrahim2023investigating}.

To address the research needs outlined above, this study investigates the use of electroencephalogram (EEG) technology to objectively assess neurophysiological responses to simulated AR-based safety warnings in roadway work zones under different workload conditions. The objectives of this study are to (1) evaluate the effectiveness of AR multi-sensory warnings in enhancing situational awareness and attention in a simulated roadway work zone under varying workload conditions, using EEG-based brain wave analysis, (2) quantify and compare the cognitive load experienced by workers under varying workload conditions when exposed to AR multi-sensory warnings, utilizing EEG-derived cognitive load metrics. (3) examine the relationship between subjective assessments of situational awareness and objective EEG-based metrics.

This study contributes to the body of knowledge by advancing our understanding of augmented reality's potential to enhance safety and efficiency in high-risk roadway work zone environments. By integrating AR with VR simulations and utilizing advanced neurophysiological measures, this research quantifies the impacts of AR multi-sensory warnings on worker situational awareness, attention, and cognitive load. It provides evidence on how AR multi-sensory warnings can impact the situational awareness and attention of workers as well as cognitive load. The use of EEG-based brain wave analysis allows for a precise, real-time assessment of how sensory inputs influence cognitive processes crucial for safety and performance. The EEG-derived metrics employed provide a deeper understanding of how AR tools can be designed to avoid cognitive overload, thereby optimizing worker performance and reducing potential errors. Another important aspect of this research is the examination of the relationship between subjective measures of situational awareness and objective EEG-based metrics. 

The findings from this study inform the design of future AR systems for safety-sensitive applications. By demonstrating the specific conditions under which AR multi-sensory warnings are most effective, this research guides developers in creating user-centered AR interfaces that adapt to varying cognitive demands. Furthermore, the methodology developed can be applied in other fields where AR might be used, such as manufacturing and military operations, thereby broadening the impact of this work. In summary, this study not only extends the theoretical frameworks of cognitive load, attention and situational awareness in the context of AR and VR but also provides practical insights that can lead to the development of more effective and safer AR applications in professional environments.

\section{Related Work}
\label{sec2}

\subsection{Enhancing Situational Awareness Using Augmented Reality}
Augmented reality has emerged as a promising technology for enhancing situational awareness across various domains. By overlaying real-time information onto users' environments, AR enables a more comprehensive understanding of surroundings and maintains situational awareness \cite{woodward2022analytic}. 
In the transportation sector, AR has been employed in driving context to improve drivers' SA in multiple scenarios, including detecting vehicles and pedestrians \cite{
phan2016enhancing, kim2022assessing}, navigation \cite{lin2011design}, identifying road signs and providing AR driving assistance to novice drivers \cite{lin2011design, rane2016virtual}. Wohn et al. \cite{kim2011effects} compared navigation using 2D maps (e.g., Google Maps) to an AR-enabled navigation and found that AR navigation led to faster route decisions at complex points, though it increased cognitive workload. In the context of worker safety, AR was used in roadway work zones \cite{sabeti2024augmented, sabeti2024mixed} for advanced warning delivery to workers. In construction industry, AR was examined for facility mangers to augment information about the facilities they maintain \cite{irizarry2013infospot} where Using AR led to lower cognitive workload and faster detection of information.  
Wallmyr et al. \cite{wallmyr2019evaluating} investigated if AR could benefit the SA of excavator operators in a simulated excavator environment with augmented icons would appear on the simulated windshield. Using AR led to lower cognitive workload and faster detection of information. 
Ramos-Hurtado et al. \cite{ramos2022proposal} investigated the deployment of AR in construction safety, proposing a methodology to enhance safety inspections through AR technology. Their study identified the key performance indicators needed for effective safety inspections and demonstrated how AR could improve situational awareness (SA) by superimposing virtual safety elements onto real construction sites.
Lukosch et al. \cite{lukosch2015providing} explored the use of augmented reality (AR) to improve situational awareness and collaboration in the security domain. Their study found that AR enhances information exchange and situational awareness, particularly for remote users, leading to better team collaboration during security operations.
In military and security applications, AR has been examined for flying Unmanned Aerial Vehicles (UAVs) \cite{ruano2017augmented}, augmenting representation of friendly or hostile forces\cite{neuhofer2012adaptive,le2010use}, navigation aids \cite{neuhofer2012adaptive,brandao2017using}, and visualization of occluded entities \cite{livingston2011user}. The industrial sector has explored the use of AR for monitoring production processes \cite{novak2014interactive} and providing maintenance \cite{aschenbrenner2016artab}. 
In the medical field, studies have looked into using AR to aid in monitoring patient information \cite{liu2010monitoring}. 
The aforemoentioned AR applications across different domains have utilized various mechanisms to augment the real world, including through smartphone apps \citep{baskar2022framework,rane2016virtual}, heads-up displays in cars \citep{
park2015augmented,phan2016enhancing}, and AR eyewear \citep{sabeti2024mixed}.

While research has demonstrated the significance of AR in enhancing situational awareness, most studies have focused on applicability and usability, with only a few systematically assessing situational awareness. Among these, most relied on subjective measures, such as post-trial ratings using the Situation Awareness Rating Technique (SART) \cite{selcon1991workload} and techniques such as the freeze-probe recall method using  Situation Awareness Global Assessment Technique (SAGAT) \cite{endsley1988design}. This highlights a gap in the current literature, where objective and comprehensive assessments of AR’s impact on situational awareness are still under explored, particularly for delivering augmented reality warnings for worker safety. 
Furthermore, some studies have shown that while AR can improve SA, it may also result in increased cognitive workload in certain applications. The performance of AR highly depends on its design; poorly designed systems in SA-critical domains could lead to severe consequences, including aircraft crashes \cite{stanton2001situational}, medical errors \cite{fore2013concept}, and worker injuries or fatalities. 

In summary, while AR has demonstrated potential for enhancing SA, there is a need for more research focused on objective measures of SA, such as wearable technology that captures brain activity data. Utilizing these objective measures in the design of AR technologies could help optimize information presentation, reducing distraction and cognitive load.

\subsection{Capturing Psyhco-physiological and Behavioral Data to Quantify Situational Awareness and Cognitive Metrics}
A review of the literature reveals a diverse array of methods utilized to evaluate and measure SA through various cognitive metrics. These methods can be classified into subjective and objective approaches. 
Subjective approaches primarily rely on inputs from participants or observers, typically collected after trial completion.  These methods often use questionnaires and include freeze probe recall, real-time probe, post trial subjective rating, and observer rating techniques \cite{salmon2009measuring}.
Some examples include SAGAT \cite{endsley1988design}, situation present assessment method (SPAM) \cite{durso1998situation}, SART \cite{selcon1991workload}, Situation Awareness Behavioral Rating Scale (SABARS) \cite{matthews2002assessing}, and Low-Event Task Subjective Situation Awareness (LETSSA) \cite{rose2018low} for SA measurement. Similar subjective approaches have also been developed for cognitive load assessment. These methods include the NASA Task Load Index (NASA-TLX) \citep{hart1988development} and Subjective Workload Assessment Technique (SWAT) \cite{tein1989subjective}. 
These frameworks have been applied in different contexts and offer several advantages, including low-cost implementation in both simulations and real-world scenarios, ease of use, and adaptability to different contexts with minimal modifications \cite{endsley2020divergence}. However, they have also faced  criticisms 
\cite{braarud2021investigating,matthews2020subjective}. Some examples of these criticisms include susceptibility to biases in post-trial feedback, impact of task interruptions on human subjects' cognitive performance, close and unduly association with workload, inconsistent interpretation strategies in different domains, and poor alignment with performance-based evaluations. A review of 37 studies by Endsley \cite{endsley2020divergence} including both objective and subjective measures of situational awareness highlighted a dissociation in outcomes of subjective and objective methodologies across a wide range of measurements. This disconnect was mainly attributed to the lack of meta-awareness about one’s own SA, poor calibration, and confounds with workload \cite{endsley2020divergence}.

In contrast, objective methodologies often involve the use of measurable and quantifiable data - such as psycho-physiological responses, performance metrics, and task analytics - to quantify an individual's SA and cognitive metrics \cite{jiang2024understanding,hu2023cognitive}. Historically, researchers have mainly relied on subjective metrics for evaluation of SA and cognitive metrics; however, the recent advancements in senors and novel algorithms have propelled studies that prioritize data-driven approaches for clinical studies in SA and cognitive assessment \cite{thombre2020sensors}. 
To reduce the influence of subjectivity in assessing SA and cognitive metrics, there is growing interest in leveraging advanced, data-driven, and sensor-based technologies. These methods aim to measure SA and cognitive load through psycho-physiological data with minimizing the reliance on subjective assumptions\cite{bracken2021can,wang2019detecting}. For this propose, a wide range of technologies and metrics has been utilized 
\cite{charles2019measuring} including: cardiovascular metrics (heart rate, heart rate variability, inter-beat interval, and normal to normal interval) \cite{solhjoo2019heart,ardecani2024assessing}; respiratory system metrics (rate, airflow, and volume of respiratory gas) \cite{grassmann2016respiratory}; skin-based measures (electrodermal activity (EDA), tissue blood volume, skin conductance level, and skin conductance response) \cite{li2021sensitivity}; blood pressure, (pressure during heart muscle contraction and relaxation) \cite{lyu2015measuring}; ocular measures (blink rate, blink duration, blink latency, and pupil size) \cite{liao2021multimodal}; and brain activity measures (electroencephalogram (EEG), functional magnetic resonance imaging (fMRI), and functional near-infrared spectroscopy (fNIRS)) \cite{broadbent2023cognitive}. Although no single method has been universally recognized as the definitive choice, there is a growing body of literature that supports the use of these measures for cognitive research \cite{charles2019measuring,ranchet2017cognitive}.

Among cognitive assessment tools, brain activity measures using EEG has emerged as a high-potential technology  
\cite{jiang2024understanding,antonenko2010using}. 
EEG offers a continuous, personalized, and real-time measure of cognitive metrics at a high temporal resolution. 
This capability is particularly useful for understanding the impact of real-time interventions, such as safety warnings, especially when general measures of cognitive load do not capture specific differences in cognitive processing \cite{antonenko2010using, alyan2023blink}. 

The outlined advantages of EEG have led to a growing interest in exploring its potential for assessing different metrics related to human cognitive performance across different disciplines, including safety in workplaces. For instance, Wang et. al \cite{wang2023identifying} proposed a wireless-based EEG device to objectively track the mental fatigue of construction workers. The EEG signals of 16 human participants were recorded during a continuous physical and cognitive task.
In another study, Kastle et. al \cite{kastle2021correlation} proposed an analytical approach for identifying EEG signals related to situational awareness across different brain regions. The proposed methodology was applied on a dataset collected from 32 participants who completed the SA test using a 32-channel EEG headset. Their results analytically pointed to a correlation between some specific frequency bands of EEG signals and SA. Ren et. al \cite{ren2023assessing} leveraged EEG to quantitatively evaluate mental workload during task switching of assembly workers. In this study, the authors used time-frequency and spectral analyses to measure and reflect task demands during the intervals between individual tasks. 

In spite of increasing interest in using EEG technology for assessing cognitive workload in workers, especially in vertical construction environments with a focus on safety, a major gap still remains in understanding how physical engagement affects workers' real-time SA, neural responses and cognitive load \cite{zhang2023digital}. While the existing body of knowledge has partially covered EEG applications in related areas, such as mental states during work at height \cite{li2024eeg}, the effects of heat stress on cognitive states \cite{shin2024impact}, and mental fatigue in construction fall hazard prevention \cite{tehrani2022assessment}, to the best of our knowledge, there has been limited exploration of the objective investigation of relationship between situational awareness, cognitive load, physical workload, and neurological responses from the perspective of EEG data in high-risk settings such as highway work zones. 

\subsection{Methods and Challenges of deriving Cognitive Insights from EEG Signals}
\label{subsec3}
Research in neuroergonomics has flourished in recent years, driven by advancements in noninvasive brain monitoring techniques that can be used to study various aspects of human behavior in relation to technology and work, including mental workload \cite{safari2024classification}, visual attention \cite{kant2020cwt}, working memory \cite{barkana2022analysis}, motor control \cite{parr2023eeg}, human-automation interaction \cite{buerkle2022adaptive}, and adaptive automation \cite{zhang2024eeg}. 

EEG signal processing methods have allowed for the real-time determination of cognitive metrics in various disciplines such as education \cite{fuentes2023low} where the authors used brainwaves to detect the level of attention of students during lecture using Power Spectral Density (PSD) of the beta band, and 
fast fourier transform using Welch's method to demonstrate significant correlation between PSD and student academic performance.
Defense, where \cite{kahraman2023neuroscience} used Fast Fourier Transform, Band power graph values, and non parametric Friedman Test to study the cognitive levels of tugboat captains during port maneuvers; transportation \cite{borghini2012assessment} where the mental fatigue and drowsiness during car driving in a simulated environment was analyzed and an EEG-based cerebral workload index was proposed for on-line assessment of drivers; aviation \cite{li2023recognising} where the cognitive workload of pilots was measured during live-flight environment using delta, theta, alpha, and beta bands using recursive feature elimination, lasso cross-validation and stacking ensemble machine learning techniques. 
Conventional linear analysis methods including Independent Component Analysis (ICA) \cite{subudhi2023automated, islam2016methods}, Fast Fourier Transform (FFT) \cite{kastle2021correlation}, Autoregressive (AR) \cite{schlogl2006analyzing}, Wavelet Transform (WT) \cite{MORETTI2003199}, and Principal Component Analysis (PCA) \cite{zhao2011automatic, wang2017monitoring} are commonly preferred techniques for analyzing EEG signals followed by nonlinear analysis methods comprising of correlation dimension (CD) \cite{pereda1998non}, Hurst exponent (H) \cite{geng2011eeg}, Higher Order Spectra (HOS) \cite{chua2011application}, Fractal dimension (FD) \cite{liu2014eeg}, recurrence plots and phase space plots \cite{7079068}.


Despite significant progress in analyzing the non-stationary nature of EEG indices to measure human performance, there are still a number of challenges, particularly due to the presence of noise and artifacts, including those caused by muscle movements, which can significantly degrade the quality of EEG signals \citep{wang2019detecting, mehmood2023deep, mir2022investigating}. Additionally, there aren't many studies that evaluate human brain signals
during multiple simultaneous cognitive tasks
that replicate real-world scenarios \cite{trejo2015eeg}.

In sumary, deriving cognitive insights from EEG signals in dynamic environments, such as workers performing tasks in VR simulations, presents  some challenges. The primary obstacle is artifact contamination due to subject movement, which is inherent in scenarios involving varying task intensities and responses to warnings. These movement artifacts, combined with the low signal-to-noise ratio typical of EEG data, make it difficult to distinguish genuine cognitive signals from noise. The non-stationary nature of EEG signals during dynamic tasks further complicates analysis. 
Real-time processing demands for safety-critical applications add another layer of complexity, requiring efficient algorithms capable of rapid data analysis. Addressing these issues necessitates advanced signal processing techniques and robust artifact removal methods.

\section{Method}

\label{sec3}

\subsection{Experimental Design and Procedure}

\label{Experimental Design}
To address the objectives of this study, we designed an experiment in a high-fidelity virtual reality setting that mirrored a real-world roadway work zone. Participants were immersed in the simulated environment and received multi-sensory augmented reality warnings while engaged in routine roadway maintenance activities with two different workload intensities. Their brain activity was collected using a wearable electroencephalography (EEG) technology. The EEG data was then analyzed to evaluate the effectiveness of the AR multi-sensory warnings in enhancing situational awareness and attention, as well as to quantify and compare the cognitive load experienced by the participants under different workload conditions. Furthermore, subjective assessments of cognitive workload were collected from the participants using NASA-TLX questionnaires. 

The University of North Carolina at Charlotte's Institutional Review Board (IRB) examined and approved the study protocol (21-0357). Initially, the study's nature, objectives, and participant roles were clearly explained to the volunteers and their informed consent was obtained. Participants were instructed to wash their hair and avoid hair products before the experiment to ensure the highest recording quality by reducing interference from skin cells, oily secretions, and sweat. Each task was designed to be completed within a maximum duration of 2 minutes per task. 
In the following, we will elaborate on the design of the experiment, the equipment used, participant characteristics, and EEG data analysis.

\begin{figure}
    \centering
    \includegraphics[width=0.98\linewidth]{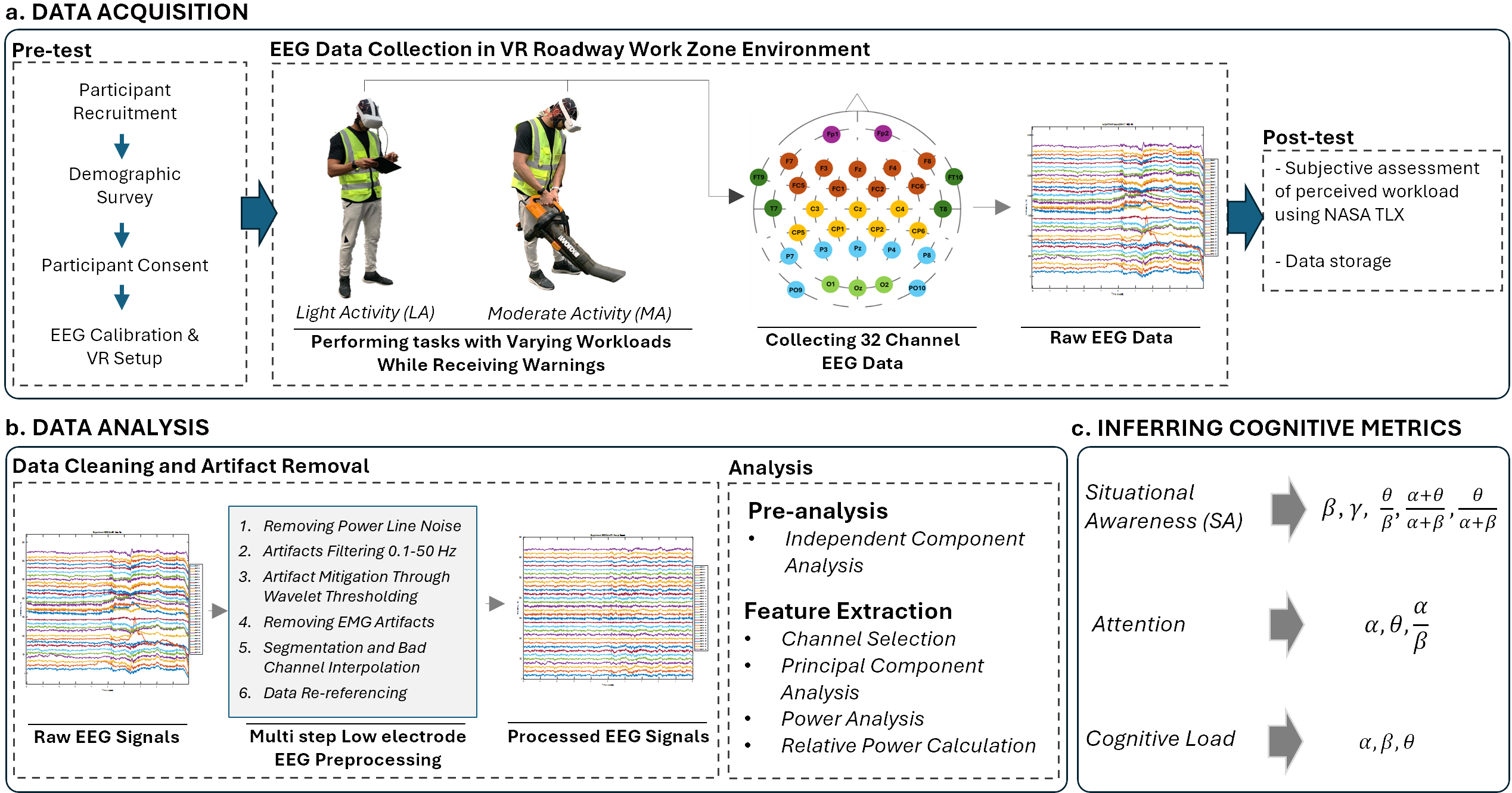}
    \captionsetup{skip=8pt}
    \caption{Overview of the main component of the methodology: (a) Data Acquisition, (b) data Analysis, (c) Inferring Cognitive Metrics}
    \label{EEG_method_Overview}
\end{figure}

\subsubsection{Task Design}

We designed two tasks categorized as light and moderate intensity to replicate typical activities performed by highway workers, thereby representing the cognitive demands associated with similar job profiles. The order of participants performing these tasks was randomized for each participant to reduce bias and the impact of the learning curve.

As shown in Figure \ref{EEG_method_Overview}, the Light Activity (LA) was simulated through an inspection task in the VR environment, where participants used a tablet to examine an obstructed drop inlet and inspect the defect by taking a picture. Drop inlets frequently become obstructed by small debris, especially in areas with high levels of precipitation leading to potential flooding, erosion of slopes, and contributing to the formation of potholes in the pavement. Therefore, making this inspection an essential and frequent roadway maintenance task. In construction, tasks that primarily involve standing, such as briefing, reading construction plans, and performing inspections, are generally classified as light activities \citep{jebelli2019application}. Such tasks typically involve minimal physical movement and are associated with low cognitive load. Moderate Activity (MA) was represented through the process of eliminating obstruction from a drop inlet using a leaf blower. In construction, MA tasks require more physical effort, such as clearing the work zone, locating tools, moving small items, and measuring and cutting materials \citep{jebelli2019application}.

\subsubsection{Immerisive Virtual Work Zone Environment}

\begin{figure}[H]
    \centering
    \includegraphics[width=0.94\linewidth]{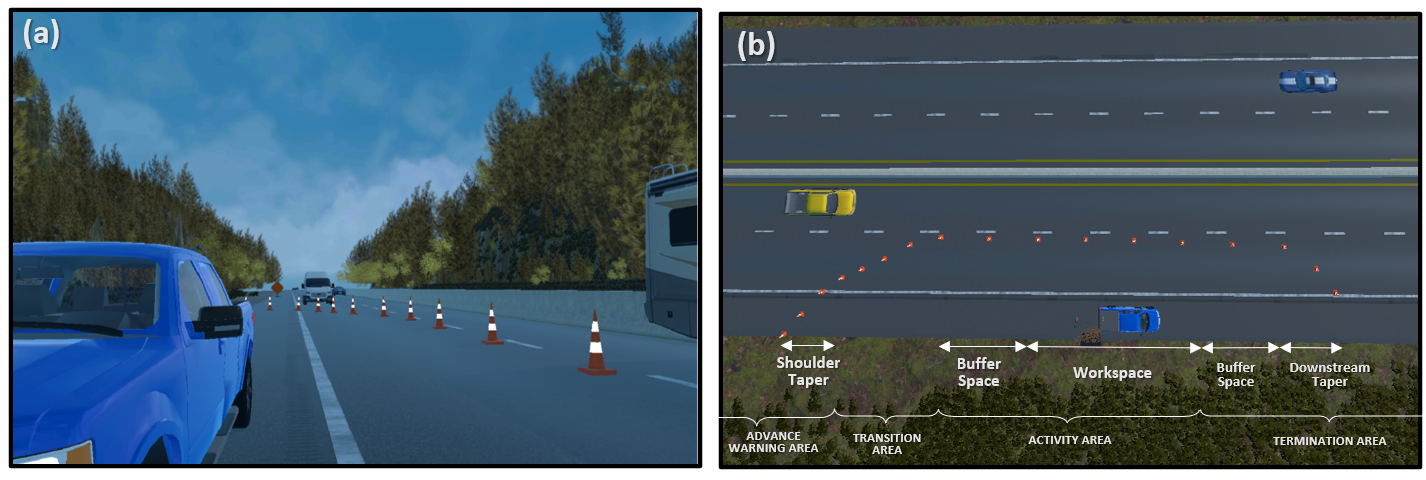}
    \captionsetup{skip=5pt}
    \caption{Features of the Developed Virtual Reality Environment. (a) Near-miss Scenario Observed from the Work Zone Perspective, Showcasing Dynamic Traffic Flow (b) Top-down View of the Work Zone Layout, Illustrating the Operational Area and Placement of Traffic Control Devices
    }
    \label{WZ_View}
\end{figure}

We designed a virtual reality environment to simulate a highway work zone based on the guidelines from the Manual on Uniform Traffic Control Devices \citep{mutcd2006manual}. This simulated work zone previously designed by the research team \citep{sabeti2024augmented} accurately reflects real-life conditions, featuring high-fidelity traffic scenarios and detailed 3D models that closely mimic a Short-duration roadway work zone. The MUTCD defines short-duration work zones as those that occupy a location up to 1 hour. They are typically used for regular operations and maintenance tasks. The work zone layout including different areas of the work zone and features of the VR environment is shown in Figure \ref{WZ_View}.

In the development of LA task in the VR simulation, a virtual model of a tablet with image-capturing capability was created within the 3D environment. Participants held a physical tablet in the real world while viewing its virtual counterpart in the VR environment. Additionally, a virtual button for capturing images was incorporated into the virtual model, and completed with a sound effect to enhance the user experience.
Building on this approach, the MA task was designed with a virtual representation of the same real-life leaf blower that participants physically held and performed the activity in the VR simulation. To enhance realism, the audio effect of the leaf blower was replicated in the VR environment, and the simulated environment included collision detection capabilities to simulate the removal of obstructions from the drop inlet. Examples of these interactions and the developed activities are illustrated in Figure \ref{Warnings_VR}

\subsubsection{Safety Warnings}

\begin{figure}[H]
    \centering
    \includegraphics[width=0.98\linewidth]{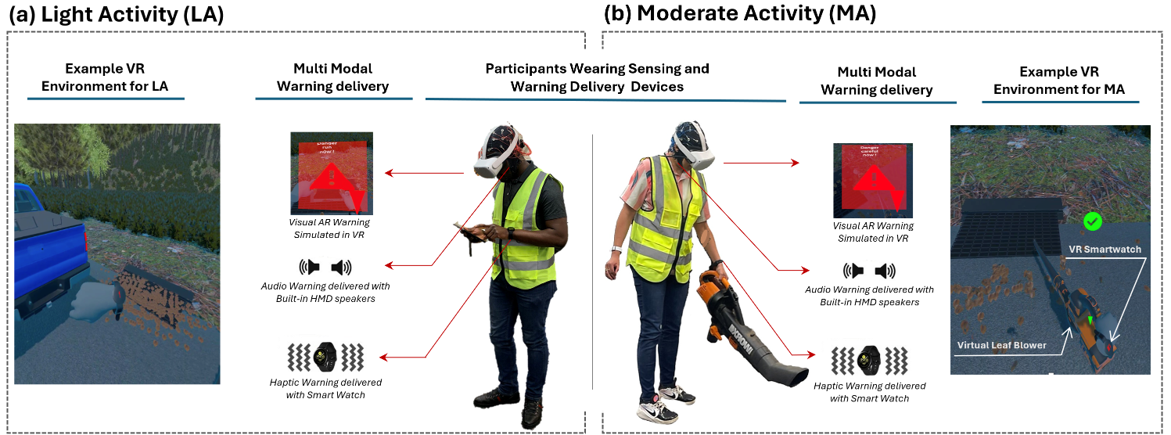}
    \captionsetup{skip=8pt}
    \caption{Features of the Developed Multi Modal Warning delivery System and Immersive Virtual Reality Environment for (a) Light activity and (b) Moderate Activity 
    }
    \label{Warnings_VR}
\end{figure}

During the experiments, participants received three identical safety warnings. These warnings were designed to be multimodal, simultaneously delivering haptic, audio, and visual cues to the users while performing tasks. The warnings were designed to activate at consistent intervals within a 1-minute scenario (25, 45, and 55 seconds). The purpose of the study was to evaluate the augmented reality safety warning technology developed by the authors \citep{sabeti2024augmented,sabeti2021toward,sabeti2022toward}.  To achieve this, the visual cue was delivered through a simulated augmented reality display within the virtual reality environment, shown in Figure \ref{Warnings_VR}(a). Haptic feedback was delivered from the smartwatch's API. In the VR simulation, participants could view a virtual smartwatch while simultaneously wearing a physical smartwatch on their wrist as shown in Figure \ref{Warnings_VR}(b).  The auditory warning consisted of a high-pitched beep at 44,100 Hz for 0.2 milliseconds and was delivered through the built-in speakers of a virtual reality headset.  These warning patterns were developed using the framework created by our team \citep{sabeti2024augmented}. 

\subsection{Experimental Equipment and Setup}

The immersive virtual reality environment was developed using the Unity 3D game engine \citep{unity} and delivered to participants through the Oculus Quest 2 headset\citep{quest2}. Augmented Reality warnings were presented through a simulated AR display within the VR setup. Participants used an Amazon Fire Tablet \citep{Amazon2024} during the LA experiment and a WORX WG509 12Amp leaf blower \citep{WORX2023} during the MA experiment. As previously discussed virtual counterparts of these tools were also simulated in the VR environment.

The haptic mode of the warnings during the experiment was delivered using a Samsung Galaxy Smart Watch \citep{SamsungElectronicsAmerica2024} facilitated by the Tizen Native framework \citep{tizen} with a predefined pattern from the smartwatch's API.

\begin{figure}[H]
    \centering
    \includegraphics[width=0.97\textwidth]{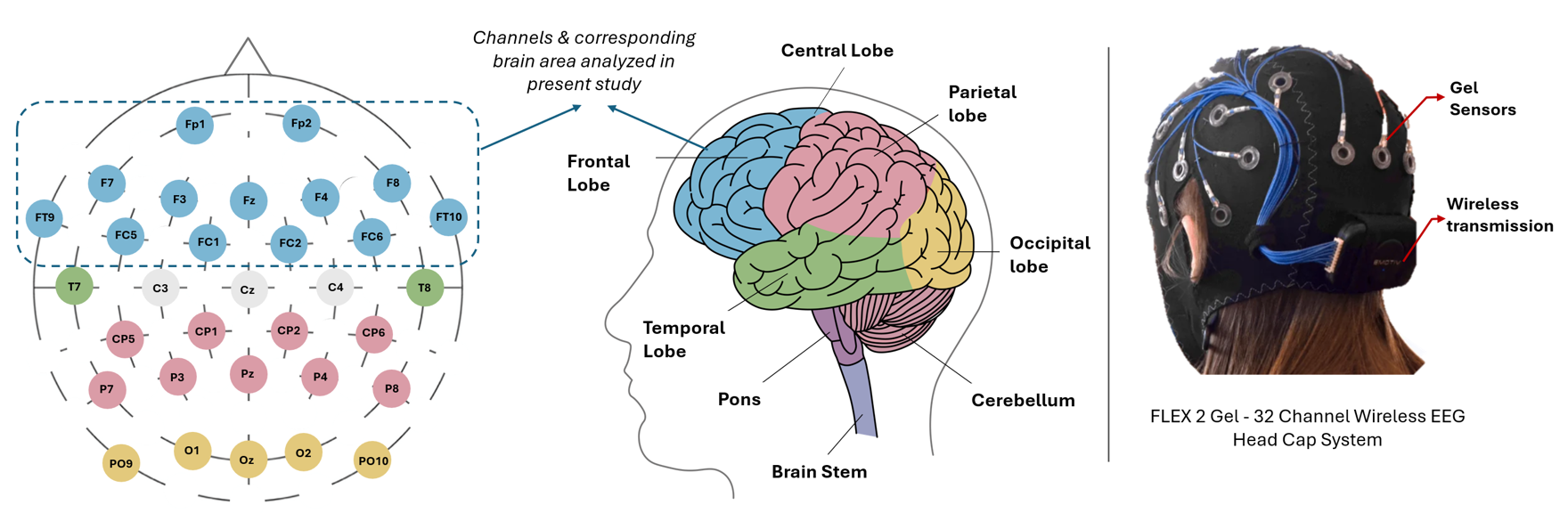}
    \captionsetup{skip=10pt}
    \caption{Spatial Configuration of EEG Electrodes, Targeted Brain Areas, and Utilized EEG Device  
    }
    \label{Cerebral-Cortex}
\end{figure}

To collect the brain activity data, we utilized Emotive's Flex 2 gel product \citep{EMOTIV2024}, a 32-channel wireless EEG Head Cap System. The device includes 32 recording electrodes, with 2 reference electrodes fitted on the scalp, and utilizes gel sensors for optimal conductance to ensure high signal quality with minimal impedance. It uses a 10-20 sensor placement to capture detailed EEG data at a sampling rate of 128 Hz and is transmitted via wireless Bluetooth technology to transmit precise contextual data to a host machine, where all the data is stored. Figure \ref{Cerebral-Cortex} demonstrates the EEG channel locations and channels used in this study.

\subsection{Participants}

We initially recruited 20 participants, however, EEG data from 3 participants were not properly recorded with significant missing data. Consequently, the final participant count was 17, consisting of 13 males and 4 females, with an average age of 27.12 years (SD = 6.82). of 17 participants, 12 had an average work experience of 3.92 years (SD = 4.78) and 5  had no prior work experience.

\subsection{EEG Data Processing and Analysis}

We developed a three-step data analysis framework designed to process raw collected EEG data and extract features that characterize situational awareness, attention, and cognitive load from the EEG signals. These steps include data cleaning and artifact removal, data analysis, and the inference of cognitive metrics such as situational awareness, attention, and cognitive workload.

\subsubsection{Data Cleaning and Artifact Removal}

The purpose of data pre-processing is to clean and prepare the raw EEG signals through several key activities—filtering, artifact removal, and re-referencing—to ensure accuracy and reliability in subsequent analyses. To mitigate noise from the EEG recordings, we used the HAPPILEE pipeline \citep{lopez2022happilee} specifically customized to address the challenges associated with lower-density EEG recordings typically using 1 to approximately 32 electrodes. This methodology has demonstrated effectiveness in identifying and excluding artifacts while preserving the integrity of the signal \citep{lopez2022happilee}. 

All pre-processing workflows were executed using MATLAB and the EEGLAB toolbox \citep{delorme2004eeglab}. Within our pre-processing pipeline, as shown in Figure \ref{EEG_method_Overview}(d), we implemented the following steps:

\begin{enumerate}
    \item \textit{Removing Power Line Noise Artifacts:} We used CleanLine \citep{cleanline} with default parameters to eliminate 60Hz power line noise artifacts, particularly the 60 Hz electrical interference commonly found in EEG recordings. CleanLine is an EEGLab plugin that uses a multi-tapering regression approach to eliminate the 60 Hz electrical noise artifact.
    \item \textit{Filtering 0.1-50 Hz to Preserve Relevant Brain Activity:} We applied a zero-phase Hamming-windowed synchronous Finite Impulse Response (FIR) bandpass filter with cutoff frequencies of 0.1 Hz and 50 Hz to the continuous data. This filter preserves relevant brain activity (0.1-50 Hz) and removes low-frequency artifacts below 0.1 Hz, such as slow drift in the signal caused by sweating or small movements of the electrodes. It also attenuates high-frequency noise above 50 Hz, which can come from muscle activity (EMG) or electrical interference from nearby equipment.
    \item \textit{Artifact Mitigation Through Wavelet Thresholding}: We employed an empirical Bayesian threshold that is soft and level-dependent for wavelets, specifically optimized for EEG data to mitigate artifacts from muscle activity (EMG), eye movements (EOG), and cardiac signals (ECG) \citep{lopez2023garments}.
    \item \textit{Removing EMG Artifacts}: After wavelet thresholding, still a high level of electromyographic (EMG) artifacts persisted in the data. EMG artifacts are electrical signals generated by muscle activity, which can significantly contaminate EEG recordings. These artifacts are particularly problematic because they can occur in frequency ranges that overlap with brain activity of interest, potentially leading to misinterpretation of the data. To address this issue, we employed MuscIL (Muscle ICLabel), an automatic EEG independent component classifier plugin within EEGLAB. MuscIL uses machine learning algorithms to classify independent components of the EEG signal into different categories, including brain activity, eye movements, muscle activity, and other artifacts. Components labeled as muscle activity with a probability threshold of at least 25\% were identified for removal. This threshold was chosen to balance between removing muscle artifacts and preserving genuine brain activity.
    \item \textit{Segmentation and Bad Channel Interpolation:} Next, we divided the continuous EEG data into smaller, equal-length segments or epochs. In this case, the data was segmented into 2-second epochs. This segmentation allows for more focused analysis of specific time periods and makes it easier to identify and handle artifacts or noise that may be present in certain parts of the recording. After segmentation, the quality of each channel within each epoch was assessed. Channels are considered "noisy" or "bad" if their signal falls below -150 µV or above +150 µV. When a bad channel was identified, it was interpolated, which means its data was estimated based on the surrounding good channels.
    This process helps to preserve the overall spatial information of the EEG while removing unreliable data.
    \item \textit{Data Re-referencing:} Finally, all channels were re-referenced to the average of all 32 scalp electrodes. The naming and spatial location of these electrodes are illustrated in Figure \ref{Cerebral-Cortex}.
\end{enumerate}

\subsubsection{Data Analysis}

After data cleaning and artifact removal steps, where we pre-processed the raw EEG data, our methodology includes a pre-analysis step to further adjust and fine-tune the data, followed by a detailed feature extraction process.

\textit{Pre-analysis:} In our pre-analysis of the data, we utilized Independent Component Analysis (ICA) to further remove any remaining artifacts. This last attempt to minimize any remaining artifacts was also recommended by the guidelines of HAPPILEE \citep{lopez2022happilee}, the process we adapted for this study. ICA is a computational method for separating a multivariate signal into additive, independent components. This step was important for isolating and removing artifacts that may not have been fully addressed during the initial pre-processing phase.
\textit{Feature Extraction:} After adjusting the data through pre-processing and artifact removal, we implemented a multi-step feature extraction process to derive meaningful information from the EEG signals. This process is crucial for transforming the cleaned EEG data into a format suitable for advanced analysis and interpretation. Our feature extraction approach consisted of the following key steps:

\begin{enumerate}
    \item \textit{Channel Selection:} We focused only on channels located in the prefrontal cortex and frontal lobe were selected for further analysis, specifically: Fz, Fp1, F7, F3, FC1, FC5, FT9, FT10, FC6, FC2, F4, F8, and Fp2. This selection was based on the significance of the frontal lobe in cognitive workload analysis. The frontal lobe is primarily associated with higher cognitive functions such as problem-solving, memory, language, judgment, and motor function. Recent research, such as López et al. \citep{lopez2023garments} and Tsolisou et al. \citep{tsolisou2023eeg}, has also emphasized the importance of focusing on the frontal lobe due to its role in executive functions and decision-making processes. By maintaining consistency in the regions of the brain being analyzed, we ensure that the data reflects the specific cognitive processes under investigation, leading to more reliable and interpretable results.
    \item \textit{Principal Component Analysis (PCA):} We utilized PCA to compute a linear combination of the underlying source signals for the EEG channels. 
    Since using all channels can be redundant and slow down processing efficiency, PCA helps reduce the dimensionality of the data, which enhances the processing speed of the algorithms and eliminates redundancy, as demonstrated in previous studies \citep{kastle2021correlation, wang2019detecting}.
    \item \textit{Power Analysis:} Using time-frequency analysis, we detected oscillations in EEG records of participants with power estimation techniques across frequencies ranging from 0.1 to 50 Hz. Time-frequency analysis of EEG signals interprets neural oscillations in terms of frequency, power, and phase. This analysis reflects fundamental neural mechanisms across spatial and temporal scales, allowing differentiation of mental states by detecting characteristics of oscillations such as magnitude, duration, and intensity spikes. This was accomplished using a 0.0625-second Hamming window, which was zero-padded to 256 points. For this purpose, we utilized Welch's method, a well-established technique for estimating the power spectral density (PSD) of signals, commonly used in signal processing and particularly in the analysis of time-series data such as EEG signals \citep{chiu2023quantifying}. 
    This method allowed us to extract power spectral densities and filter out specific frequency bands of interest: delta (1-4 Hz), theta (4-8 Hz), alpha (8-12 Hz), beta (12-30 Hz), and gamma (30-50 Hz). Since delta waves are prevalent during the deep sleep stage and typically not examined for the assessment of mental effort, we excluded this frequency band from further analysis.
    \item \textit{Relative Power Calculation:} Given the variability in EEG activation among participants, in this step we calculated relative power within each frequency band, ensuring comparability and consistency in the analysis. 
    Also, EEG activation varies across participants, relative values were utilized in this study. For this purpose, the relative amount of power within each frequency band was calculated using the following formula\citep{jang2023evaluating}, exemplified here for the beta power band:

\begin{equation}
\text{R}_{\beta_i}^p= \left( \frac{\int_{f_l}^{f_h} S_x(f) \, df}{\int_{0}^{f_{\text{max}}} S_x(f) \, df} \right)_{\beta_i}^p \times 100\%
\end{equation}

where, 
${R}_{\beta_i}^p$= the relative amount of beta wave in p channel for each subject (\%),
$p$= position of each electrode in the frontal lobe,
$f_l$= frequency lowest of beta band (12Hz),
$f_h$= frequency highest of beta band (30Hz), and
$f_{max}$= frequency highest of gamma band (50Hz)

\end{enumerate}

\textit{Extracting the Time-Window of Interest:} The primary objective of analysis was to examine brain behavior following augmented reality warnings and compare cognitive metrics obtained from EEG signals before and after these warnings. To focus on the critical pre- and post-warning EEG recordings relevant to this study and to reveal the impact of the warnings, we selectively extracted the EEG records from 5-second windows before and after each warning time in every experiment. Furthermore, to establish a reference state characterizing participants' brain activity prior to receiving warnings, we utilized a Baseline Window consisting of the 5 seconds before the warning. For the Post-Warning Window, which spanned 5 seconds after the warning, this period was divided into 20 partitions, each lasting 125 milliseconds. Given the data collection frequency of 128 Hz, each 125-millisecond segment contained 32 EEG records. These segments were then aggregated for subsequent feature extraction.

\subsubsection{Inferring Cognitive Metrics of Situational Awareness, Attention, and Cognitive Workload}

After obtaining the frequency bands of brain activity from the data analysis step, we focused on inferring three cognitive measures: situational awareness, attention, and cognitive load experienced by workers across different experimental scenarios. As shown in Figure \ref{EEG_method_Overview} (c), these cognitive measures are linked to one or a combination of frequency bands. Building on previous research, we incorporated multiple approaches and evaluated them, as detailed in Table \ref{table:EEG_metrics} and explained in the following.

\textit{Situational Awareness:} Investigating how EEG signals can accurately portray SA remains a dynamic and emerging area of research. 
Following a similar approach to Kastle et al. (2021) \citep{kastle2021correlation}, our study will focus on analyzing the beta (12-30 Hz) and gamma (30-45 Hz) bands in relation to SA as main brain activity's metrics. Additionally, we will examine combined EEG metrics such as $\frac{\theta}{\beta}$, $\frac{\alpha+\theta}{\alpha+\beta}$, and $\frac{\theta}{\alpha+\beta}$. Previous studies \citep{kang2024recognizing} have found significant correlations between these combined EEG metrics and SA.

\textit{Attention:} Attention is a significant measurement for assessing human cognitive performance. The alpha EEG rhythm, within the alpha-band frequency range, has been commonly associated with attention \citep{laberge2001attention}. A decrease in the proportion of other waves, such as theta, is necessary to sustain attention \citep{norris2001effects}. Also, in cognitive neuroscience, the attention ratio ($\frac{\alpha}{\beta}$) is a valuable factor for evaluating attentional states during cognitive tasks \citep{wan2021frontal,sacchet2015attention}. 

\textit{Cognitive workload}: 
Cognitive workload refers to the number of cognitive resources required to complete a task or series of tasks, measuring the mental demands placed on an individual during a particular activity or situation. Various factors can affect cognitive workload, including the complexity of the task, the individual's level of expertise, environmental conditions, and emotional state. A high mental workload can lead to increased stress, decreased performance, and reduced overall well-being. Therefore, evaluating cognitive workload (CWL) is essential for understanding human performance and improving task design across multiple domains, such as aviation, driving, and human-computer interaction \citep{li2020understanding}.

We used both subjective and objective methods to gauge cognitive workload. Subjective assessment was performed using the NASA Task Load Index (NASA-TLX), an assessment tool designed to evaluate the workload experienced by individuals during task performance It quantifies CWL using a questionnaire that considers six factors: mental demand, physical demand, temporal demand, performance, effort, and frustration. The objectvie measurement was accomplished through EEG signal analysis.
Research has demonstrated a significant relationship between EEG signals, in particular, alpha, beta, and theta waves, in different frequency bands and CWL \citep{puma2018using, jao2020eeg}.

\begin{table}[H]
\centering
\renewcommand{\arraystretch}{1.77} 
\scriptsize
\caption{Summary of EEG Metrics for Evaluating Situational Awareness and Cognitive Measures Across Various Application Contexts}
\vspace{9 pt}
\label{table:EEG_metrics}
\begin{tabular}{lp{4.55cm}p{7.5 cm}}
\toprule
EEG Metric& Key Insights & Application Context \\ 
\midrule
\multicolumn{3}{l}{\textit{\textbf{Situational Awareness}}} \\
\midrule
\multirow{2}{*}{$\beta$} 
& - Strong correlation between $\beta$ and SA levels& 
- Real-time SA assessment to improve operator safety \citep{kastle2021correlation}, Driving performance after sleep deprivation \citep{EOH2005307} \\ 
& - Increased frontal beta activity indicates higher SA & 
- Impact of task complexity on SA in air traffic control \citep{li2021situational} \\ 
\cmidrule(lr){2-3}
\multirow{2}{*}{$\gamma$} 
& - Association between $\gamma$ rhythm and attention, working memory, decision making, \& learning activities & 
- Monitoring construction workers vigilance \& attention \citep{wang2017monitoring}, Drug-resistant epileptic patients during task execution \citep{BABILONI2016641} \\ 
& - Correlation between $\gamma$ and SA & 
- Real-time SA assessment for improving safety \citep{kastle2021correlation} \\ 
\cmidrule(lr){2-3}
$\frac{\theta}{\beta}$, $\frac{\alpha+\theta}{\alpha+\beta}$, $\frac{\theta}{\alpha+\beta}$ 
& - Combined EEG metrics in the frontal lobe correlate with SA & 
- Forklift operators' SA performing tasks like loading, driving, and unloading \citep{li2024recognizing}, Monotonous driving sessions with few road stimuli \cite{jap2009using}, Driving forklift sessions with tasks like stacking and unstacking, and road test \cite{kang2024recognizing} \\ 
\midrule
\multicolumn{3}{l}{\textit{\textbf{Attention}}} \\
\midrule
\multirow{2}{*}{$\alpha$} 
& - Alpha activity changes during working memory retention, indicating involvement in cognitive processes and attention & 
- Marine pilots constructing bridges and assessing performance limits \citep{orlandi2018measuring}, rapid visual stimulus processing \citep{klimesch2012alpha} \\ 
& - Higher attentiveness results in more desynchronized $\alpha$ activity & 
- Air Traffic Management simulation to assess air traffic controllers using the rule \& knowledge taxonomy \citep{borghini2017eeg} \\ 
\cmidrule(lr){2-3}
$\theta$ 
& - Theta band activity in the frontal cortex is linked to attention and working memory & 
- Visual or tactile stimuli tasks, memory recall \citep{jensen2007human}, Visuomotor aviation simulation \citep{ sterman1994multiband} \\ 
\cmidrule(lr){2-3}
$\frac{\alpha}{\beta}$ 
& - A feature used to determine attentiveness levels & 
- Assessing student attentiveness during the learning process \citep{liu2013recognizing} \\ 
\midrule
\multicolumn{3}{l}{\textit{\textbf{Cognitive Workload}}} \\
\midrule
$\alpha$ 
& - Linked to increased mental loads & 
- During speech \citep{abhang2016introduction}, Assesing participants' cognitive workload during gauge monitoring, tracking, letter detection, and mental arithmetic \cite{puma2018using} \\ 
\cmidrule(lr){2-3}
$\beta$ 
& - Increase in $\beta$ power spectra is associated with higher mental load and cognitive demand & 
- Assessing air traffic controllers’ mental state during a 2-hour simulated task \citep{dasari2017ica} \\ 
\cmidrule(lr){2-3}
$\theta$ 
& - Theta waves are a reliable measure of mental workload & 
- Simulated human-machine system for spacecraft air quality management with varying task difficulties \cite{yin2014identification}, 
Driving simulation with concurrent visual and auditory stimuli response tasks \cite{kong2015investigating} 
\\ 
\bottomrule
\end{tabular}
\end{table}

\section{Results}

\label{sec4}

Following the process outlined in the method section, we conducted an analysis of neural responses mainly in the frontal lobe, focusing specifically on the $\alpha$, $\beta$, $\theta$, and $\gamma$ frequency bands. 
In the following, we present the findings and discuss the neurological patterns observed for three critical measures of situational awareness, attention, and cognitive load immediately after the delivery of the safety warnings. We also compare these patterns when participants were engaged in low-intensity activity versus moderate-intensity activity.
Our findings center on a 500-millisecond (ms) time window following the warning delivery. This interval was selected based on our observation that beyond 500 ms, the metrics typically returned to within ±10\% of the baseline values, indicating a stabilization of neural responses. This time frame allows us to capture the immediate and most significant changes in brain activity in response to the safety warnings.

\subsection{Situational Awareness}

\begin{figure}[H]
    \centering
    \begin{subfigure}[b]{0.9\textwidth}
        \centering
        \includegraphics[width=\textwidth]{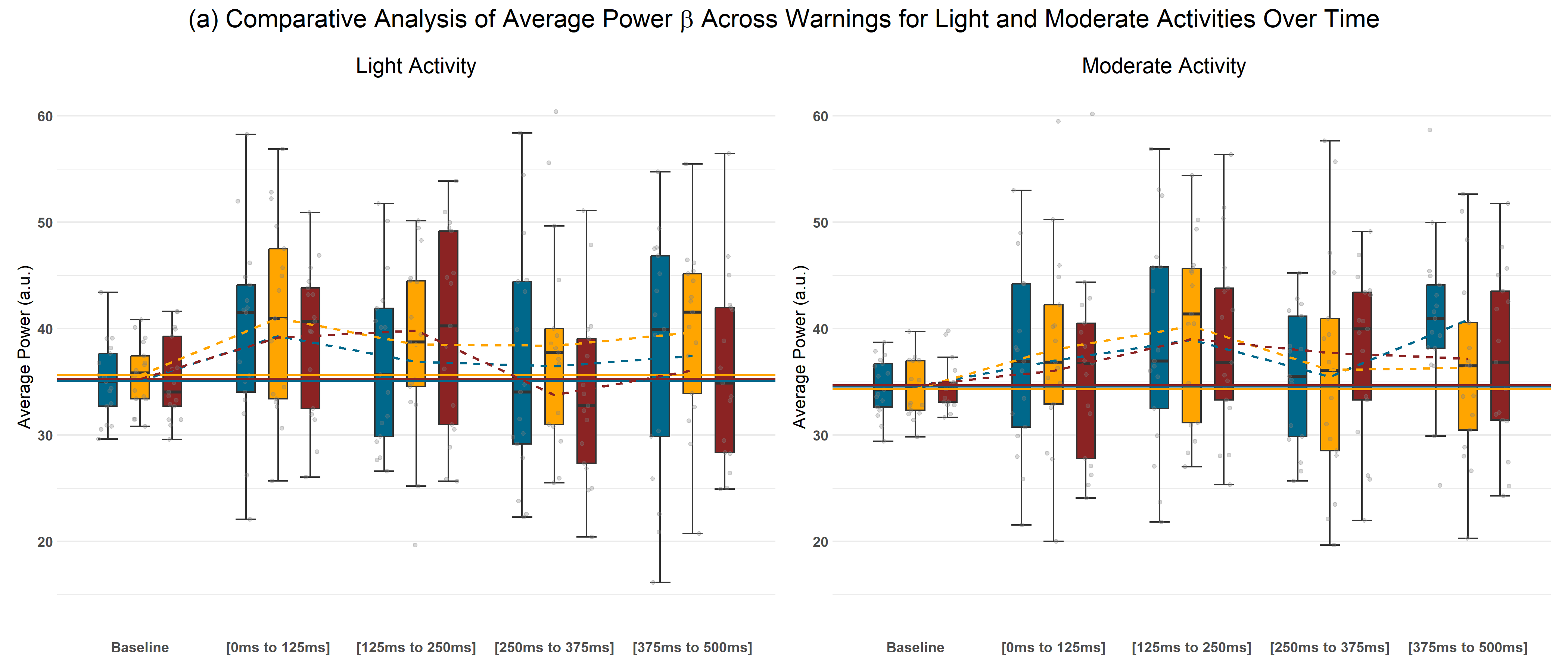}
        \label{fig:beta}
    \end{subfigure}
    
    \vspace{-1em} 

    \begin{subfigure}[b]{0.9\textwidth}
        \centering
        \includegraphics[width=\textwidth]{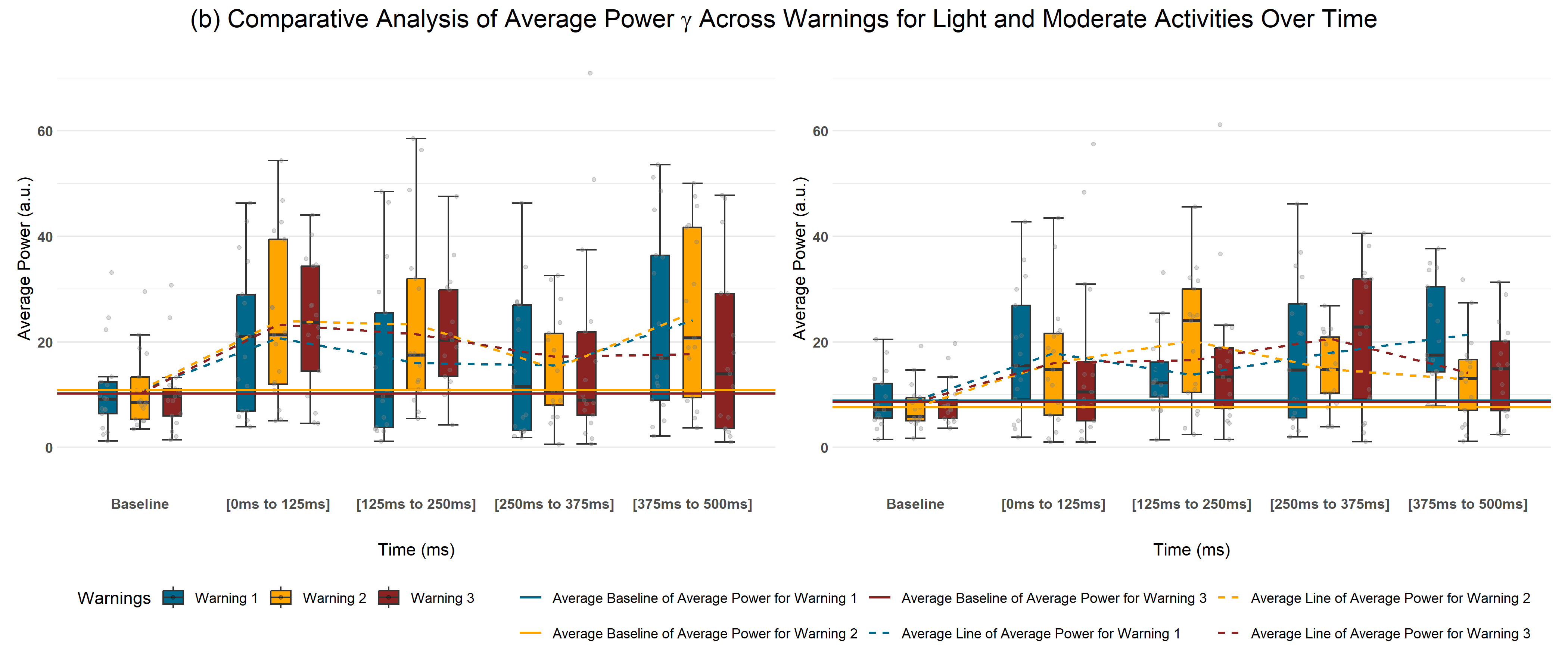}
        \label{fig:gamma}
    \end{subfigure}
    \vspace{- 15 pt}
    \caption{Comparison of Beta ($\beta$) and Gamma ($\gamma$) Waves in Light and Moderate Activity}
    \label{fig:combined_SA}
\end{figure}

\begin{figure}[H]
    \centering

    \begin{subfigure}[b]{0.9\textwidth}
        \centering
        \includegraphics[width=\textwidth]{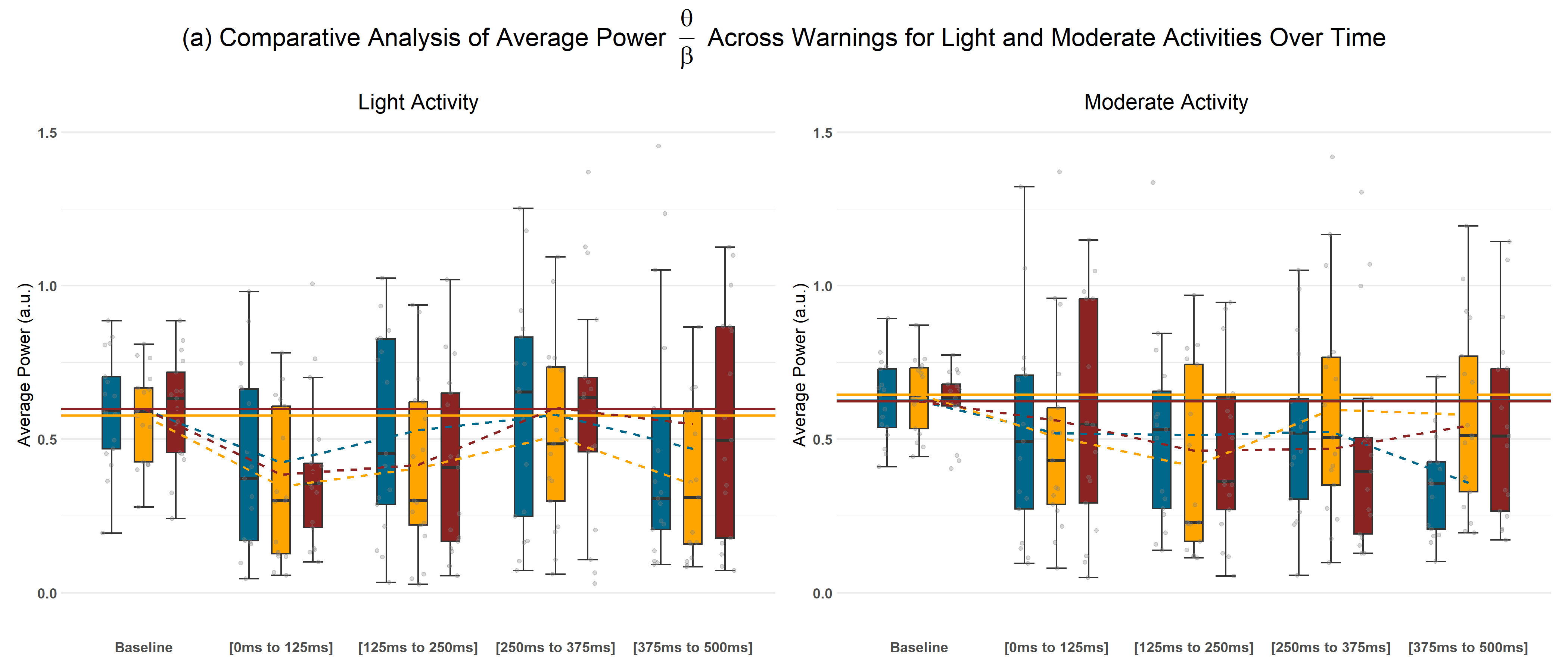}
        \label{fig:ThetaToBeta}
    \end{subfigure}
    
    \vspace{-1em} 

    \begin{subfigure}[b]{0.97\textwidth}
        \centering
        \includegraphics[width=\textwidth]{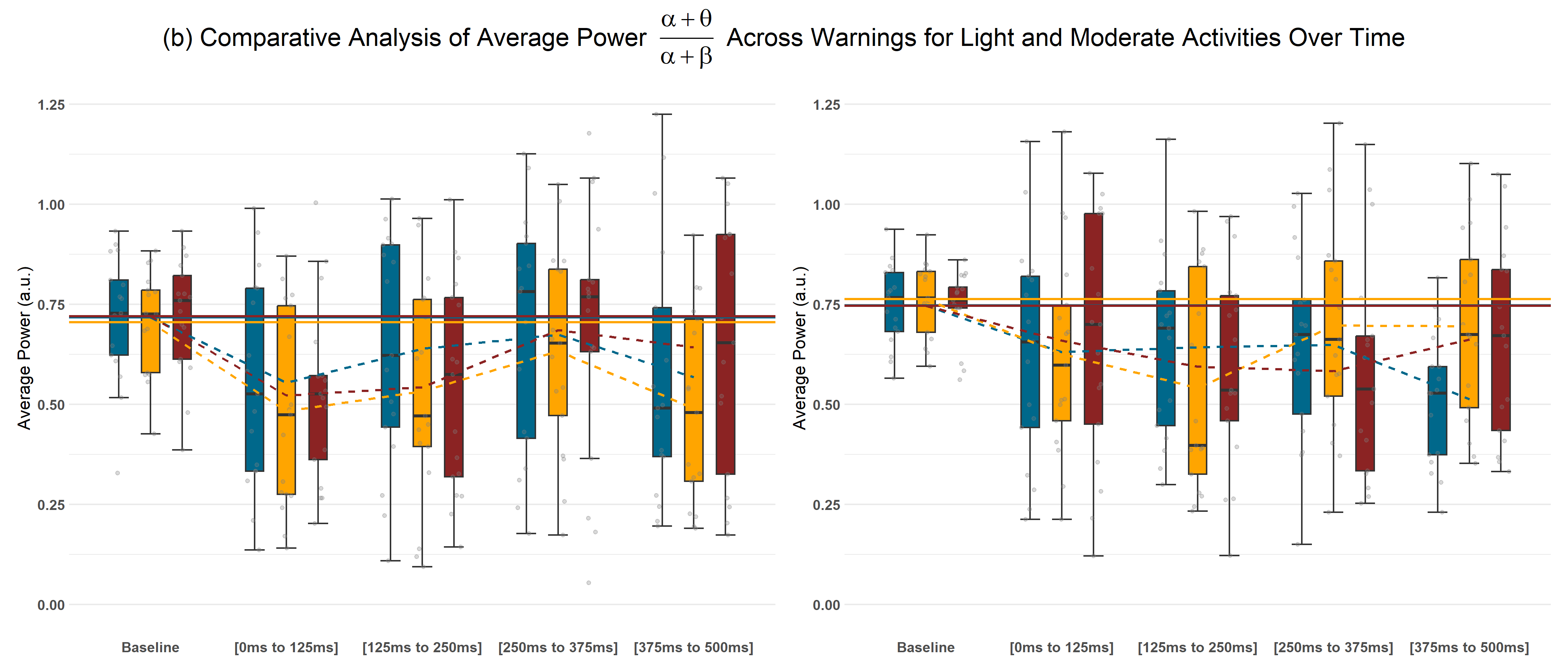}
        \label{fig:(Alpha+Theta)To(Alpha+Beta)}
    \end{subfigure}
    
    \vspace{-1em} 

    \begin{subfigure}[b]{0.94\textwidth}
        \centering
        \includegraphics[width=\textwidth]{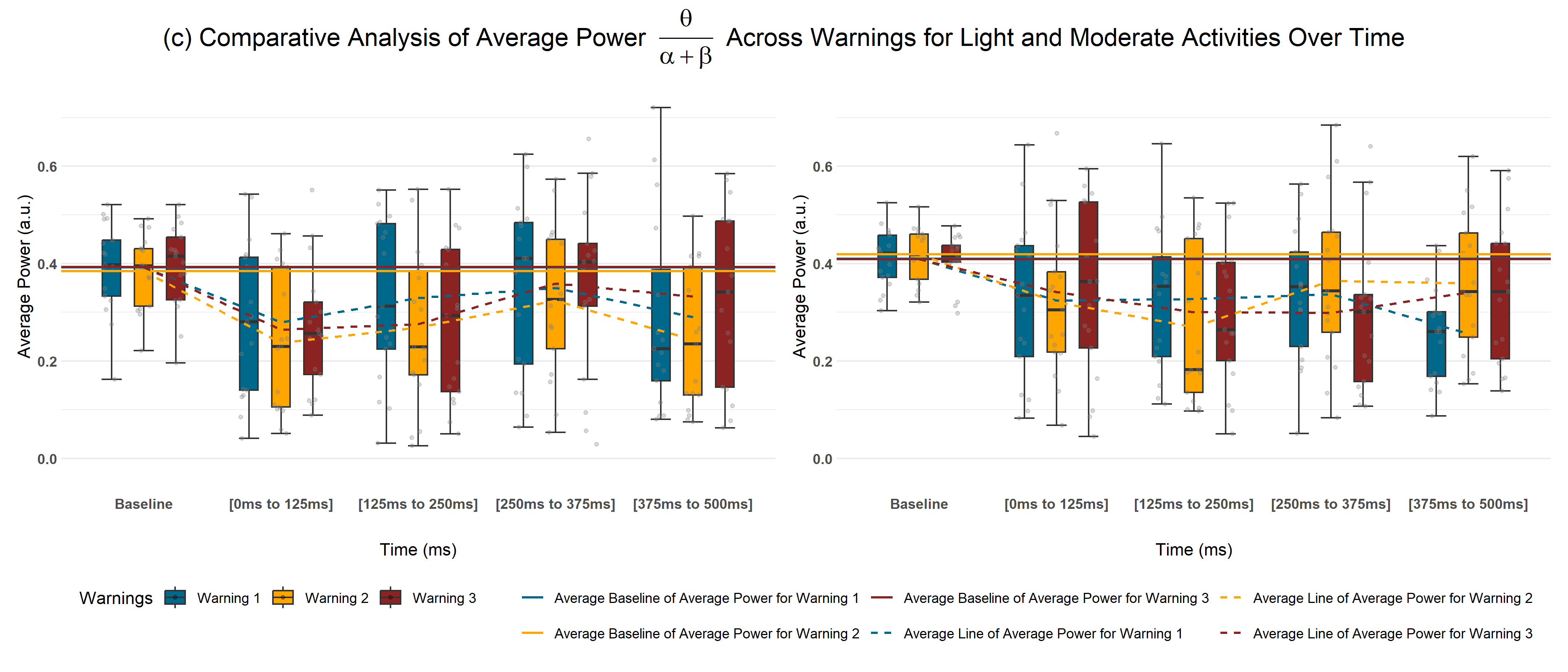}
        \label{fig:ThetaTo(Alpha+Beta)}
    \end{subfigure}
    \captionsetup{skip=-15pt} 
    \caption{Comparative Analysis of Combined EEG Ratio in Light and Moderate Activity}
    \vspace{-5 pt}
    \label{fig:combined_ratio_SA}
\end{figure}

To assess situational awareness after receiving the safety warnings, we utilized several key neurological indicators: beta power, gamma power, theta-beta ratio, theta/(alpha+beta) ratio, and (alpha+theta)/(alpha+beta) ratio. 
These metrics have been established as measures of situational awareness in various cognitive tasks. Figures \ref{fig:combined_SA} and \ref{fig:combined_ratio_SA} presents our findings for all these metrics across three warnings in two experimental conditions of low-intensity activity (LA) and moderate-intensity activity (MA). This visualization allows for a comparison of how different neural metrics of situational awareness evolve in response to warnings under different levels of physical engagement.

As shown in  Figures \ref{fig:combined_SA} and \ref{fig:combined_ratio_SA}, our analysis reveals the following patterns across the measure of situational awareness:
\begin{itemize}
    \item Beta ($\beta$) and Gamma ($\gamma$) Power: Both LA and MA conditions showed an increase in $\beta$ and $\gamma$ power immediately following each warning, indicating heightened alertness and information processing. However, the increase was more pronounced in the LA condition, suggesting that lower physical engagement may allow for more robust situational awareness responses. Moreover, as evidenced by the beta activity, the peak value of average power in beta waves was observed within the first 125 milliseconds (ms) during the LA. In contrast, during the MA, this peak consistently occurred between 125 to 250 ms for all three warnings. This temporal shift in peak beta activity suggests a differential neural response pattern based on the level of physical engagement. The earlier peak in the LA potentially reflects a greater readiness state in participants when physical demands are lower.
\end{itemize} 

\begin{itemize}
    \item Theta-Beta Ratio ($\frac{\theta}{\beta}$): This ratio indicates how much resting brainwave activity (theta) one has versus active brainwave activity (beta). A higher ratio suggests more resting activity, while a lower ratio indicates more active processing, often associated with situational monitoring and cognitive control. This ratio showed a consistent decrease across all warnings in both conditions, indicating higher situational awareness. The decline was more rapid in the LA condition and was also consistent with the peaking pattern of the beta power, where the peak for LA occurred earlier than MA.
    
    \item (Alpha+Theta)/(Alpha+Beta) Ratio ($\frac{\alpha+\theta}{\alpha+\beta}$): This metric, linked to the balance between relaxed awareness and active processing, demonstrated a sharp initial decrease followed by a gradual return to baseline. The MA condition showed a more prolonged depression of this ratio, and later peaking compared to LA.
    
    \item Theta/(Alpha+Beta) Ratio ($\frac{\theta}{\alpha+\beta}$): This ratio, indicative of cognitive workload in maintaining situational awareness, exhibited an initial decrease followed by a gradual return to baseline in both conditions. The decrease was more pronounced in the LA condition, possibly reflecting a more intense cognitive response to the warnings.
\end{itemize}

Across all metrics,  we observed that the warnings effectively triggered neurological responses associated with situational awareness as soon as 125 ms after the delivery of the warning. However, the peaking of situational awareness occurred earlier in the LA condition. These findings support that varying levels of physical activity influence the neurological underpinnings of situational awareness in response to safety warnings. The results underscore the importance of considering the physical context when designing and implementing warning systems to ensure optimal situational awareness across different activity levels.

\subsection{Attention}

\begin{figure}[H]
    \centering
    \begin{subfigure}[b]{0.9\textwidth}
        \centering
        \includegraphics[width=\textwidth]{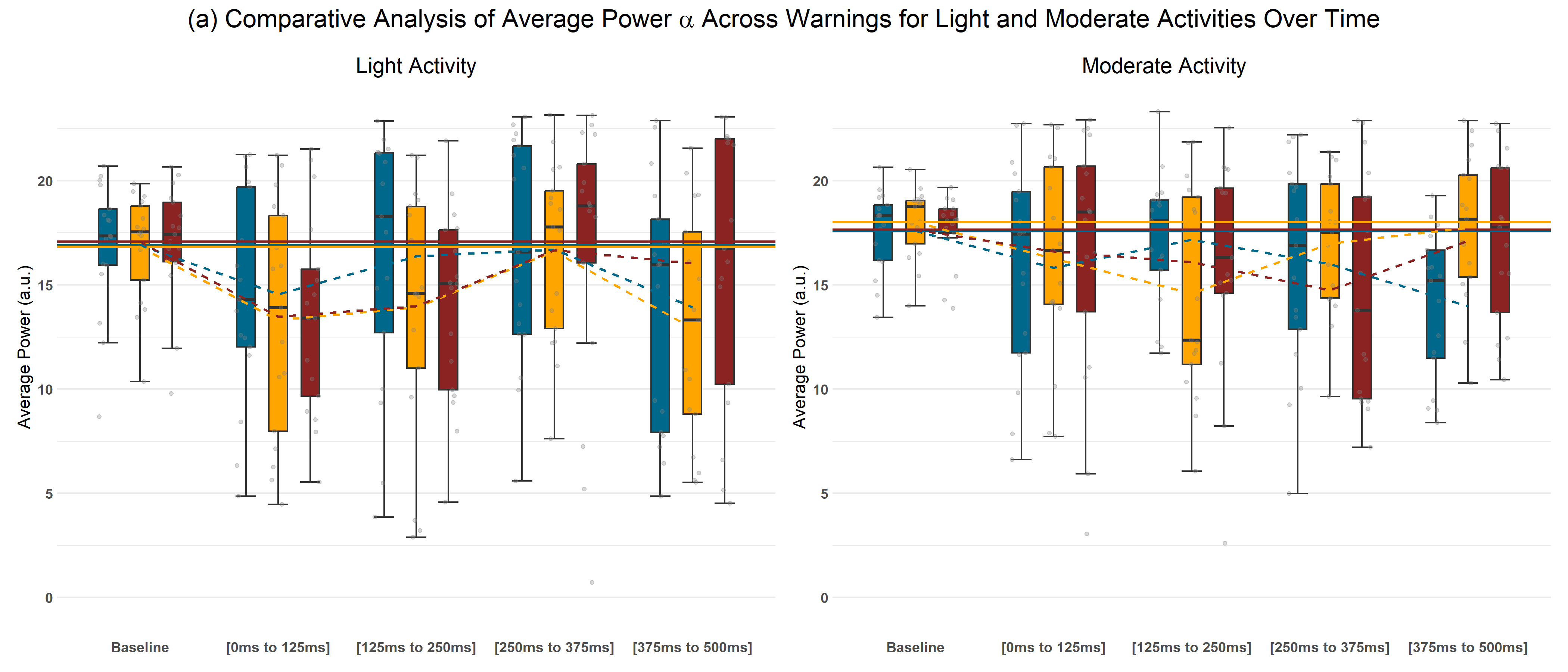}
        \label{fig:alpha}
    \end{subfigure}
    
    \vspace{-1em} 

    \begin{subfigure}[b]{0.9\textwidth}
        \centering
        \includegraphics[width=\textwidth]{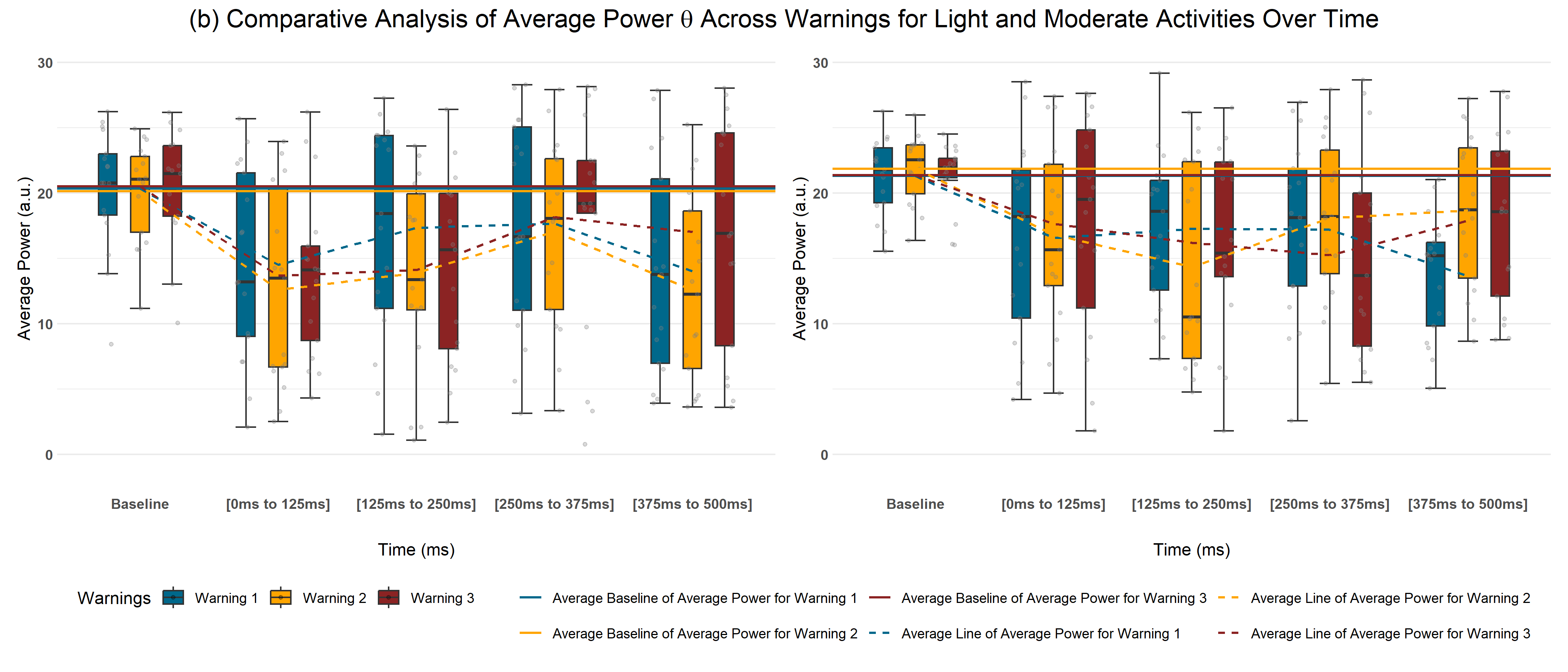}
        \label{fig:theta}
    \end{subfigure}
    \captionsetup{skip=-15pt}  
    \caption{Comparison of Alpha ($\alpha$) and Theta ($\theta$) Waves in Light and Moderate Activity}
    \label{fig:combined_Attention}
\end{figure}

\begin{figure}[H]
    \centering
    \includegraphics[width=0.9\textwidth]{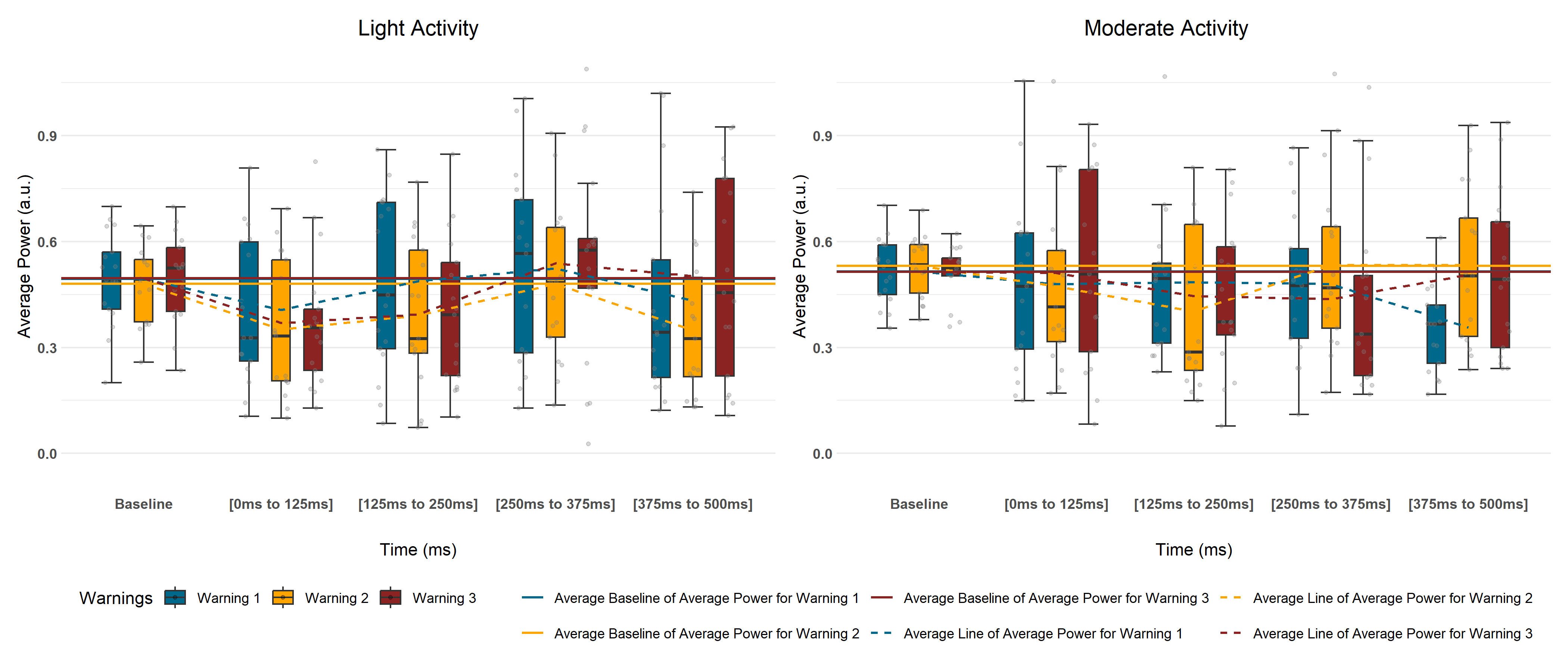}
    \caption{Comparison of the Combined ($\alpha$/$\beta$) Ratio in Light and Moderate Activity}
    \label{fig:combined_AR}
\end{figure}

To assess participants' attention levels in response to safety warnings, we analyzed three key neurological indicators: alpha waves in the back lobe, theta waves, and the alpha/beta ratio ($\frac{\alpha}{\beta}$). These metrics are recognized as reliable indicators of attentional processes. Figures \ref{fig:combined_Attention} and \ref{fig:combined_AR} presents our findings for these attention-related metrics across three warnings in both low-intensity activity (LA) and moderate-intensity activity (MA) conditions. As shown in Figures \ref{fig:combined_Attention} and \ref{fig:combined_AR}, our analysis reveals the following patterns:
\begin{itemize}

    \item Alpha (${\alpha}$) Power: We observed a consistent decrease in alpha power immediately following each warning in both LA and MA conditions. This alpha suppression is typically associated with increased attention and visual processing. Interestingly, the suppression was more pronounced in the LA condition, suggesting that lower physical engagement may allow for more robust attentional responses to stimuli.

    \item Theta (${\theta}$) Power: Theta power showed a decrease following each warning, particularly in the frontal regions. A decrease in theta waves, particularly when accompanied by an increase in beta waves, often suggests a shift from a relaxed state to a more attentive and alert state. This decrease in theta could also suggest a shift from internal focus to external, task-oriented attention. The peak value of  $\theta$ activity in the frontal lobe was observed within the first 125 milliseconds (ms) during the LA task. In contrast, in the MA task, this peak occurred at varying intervals for each warning: within the first 125 ms for the first warning, between 125 to 250 ms for the second warning, and between 250 to 375 ms for the third warning.
    \item Alpha/Beta Ratio ($\frac{\alpha}{\beta}$): This ratio, often used as an index of attentional control, demonstrated a decrease following each warning in both conditions. However, the decrease was more rapid and pronounced in the LA condition (as shown in Figure \ref{fig:combined_AR}), suggesting quicker attentional shifts. In the MA condition, the ratio showed a more gradual decline and slower recovery, indicating a potentially more sustained but less intense attentional response. The absolute peak value of the mean line of this indicator was observed within the first 125 milliseconds (ms) during the LA task. In contrast, in the MA task, this absolute peak occurred at varying intervals for each warning: within the first 125 ms for the first warning, between 125 to 250 ms for the second warning, and between 250 to 375 ms for the third warning and this trend is similar within the $\alpha$ activity of the cerebral cortex. 
\end{itemize}

Across all metrics, we observed that the timing of peak responses differed between conditions. In the LA condition, peak attentional responses (as indicated by maximum alpha suppression and theta decrease) occurred within the first 125 milliseconds post-warning. In contrast, these peaks were delayed to 125-250 ms in the MA condition. This temporal shift aligns with our observations in the situational awareness metrics, further supporting the hypothesis that increased physical activity may introduce a slight lag in cognitive responses to external stimuli. The MA condition showed more gradual changes and longer recovery times across all attention metrics compared to the LA condition.

\subsection{Comparison of Subjective Cognitive Workload Measures and EEG Frequency Bands}

Apart from the mental workload demands on workers during task execution, mental workload also encompasses other components that contribute to the overall workload at the job site.
The results presented in Table \ref{table:nasa_tlx} provide a comprehensive analysis of the correlation between cognitive workload as subjectively measured by the NASA Task Load Index (NASA-TLX) and brain frequency bands (alpha, beta, and theta) across different activity levels while receiving safety warnings. Overall, the data reveals no consistent or strong correlations between the subjective measures of cognitive workload and the EEG-based frequency band data. The correlation coefficients vary widely across different dimensions of the NASA-TLX, and activity levels (light and moderate).

For the Mental Demand dimension, correlation coefficients range from weak negative to weak positive correlations across all frequency bands and warnings. Similar inconsistent patterns are observed for Physical Demand, Temporal Demand, Performance, Effort, and Frustration dimensions. Some isolated instances of moderate correlations are observed. However, these isolated correlations do not form a consistent pattern across conditions or frequency bands, suggesting they may be chance occurrences rather than indicative of a robust relationship.
The lack of significant and consistent correlations between subjective cognitive workload measures and EEG frequency bands suggests that these two methods may be capturing different aspects of cognitive workload. 

Additionally, creating standard, generic indicators for certain construction activities is difficult due to individual variations. These differences result in varied signal intensities and patterns, even when performing the same task. The unique ways individuals perform tasks lead to inconsistent data, making it challenging to develop uniform indicators.

\begin{table}[H]
\centering
\footnotesize
\caption{Comparison of NASA TLX Measures across Frequency Power Bands}
\label{table:nasa_tlx}
\begin{tabular}{lcccccc}
\toprule
\diagbox{NASA-TLX}{Frequency\\Band} & \multicolumn{2}{c}{$\alpha$} & \multicolumn{2}{c}{$\beta$} & \multicolumn{2}{c}{$\theta$} \\
\cmidrule(lr){2-3} \cmidrule(lr){4-5} \cmidrule(lr){6-7}
 & LA & MA & LA & MA & LA & MA \\
\midrule
CWL             & 0.02 & 0.41  & -0.03 & -0.24 & 0.00 & 0.37 \\
Mental Demand   & 0.12 & 0.11  & -0.41 & -0.06 & 0.24 & 0.09 \\
Physical Demand & -0.35 & -0.34 & 0.06 & 0.50  & -0.27 & -0.41 \\
Temporal Demand & 0.16 & 0.41  & 0.09 & -0.16 & 0.91 & 0.34 \\
Performance     & 0.13 & 0.17  & 0.05 & -0.33 & 0.01 & 0.25 \\
Effort          & -0.01 & 0.52  & 0.06 & -0.32 & -0.02 & 0.45 \\
Frustration     & -0.20 & 0.32  & 0.28 & -0.17 & -0.27 & 0.28 \\

\bottomrule
\end{tabular}
\end{table}

\subsection{Influence of Physical Activities and Warning Scenarios on Brain Waves}
\begin{figure}[H]
    \centering
    \includegraphics[width=0.99\textwidth]{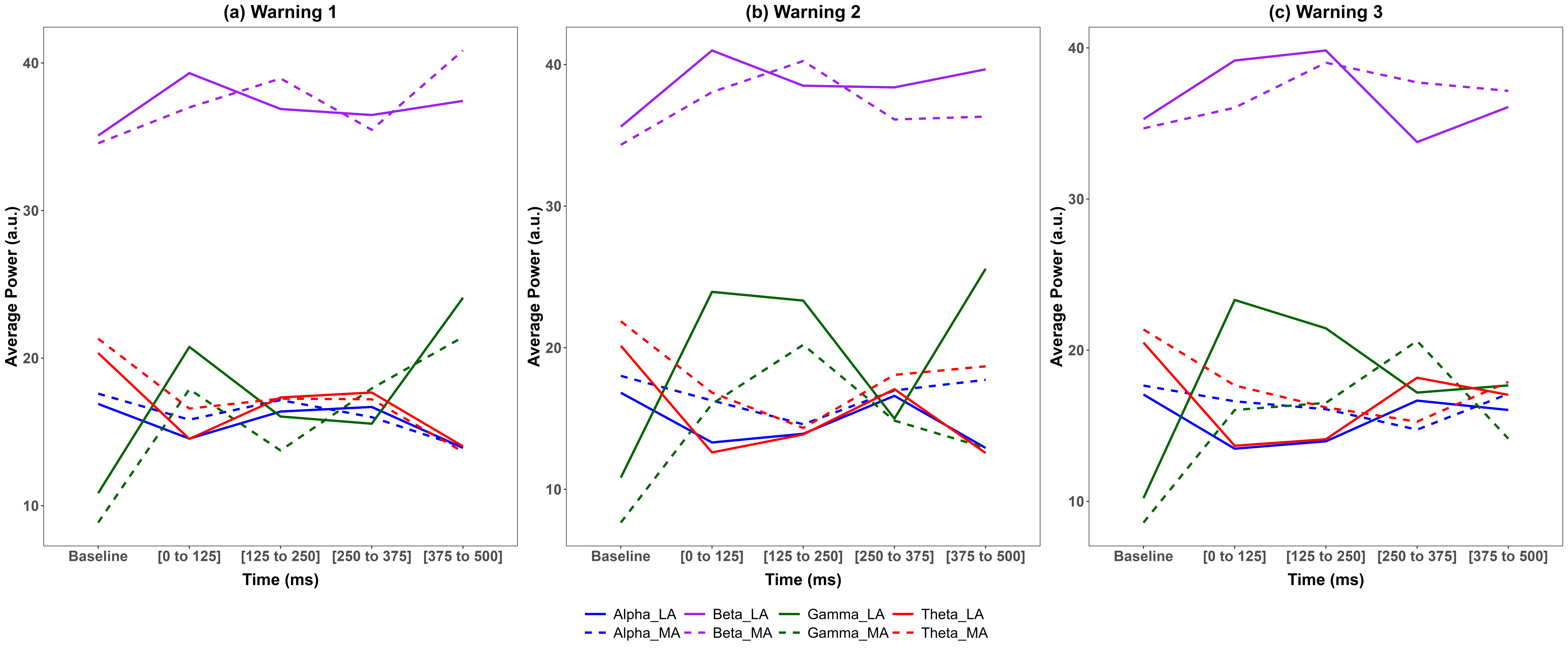}
    \caption{Comparison of warnings through the average power of all participants in light and moderate activity}
    \label{fig:compare_warnings}
\end{figure}

In our study, illustrated in Figure \ref{fig:compare_warnings}, we observed that beta waves, heightened beta waves is linked to states of focused attention, problem-solving, and deep concentration, during moderate tasks initially started at a lower level and increased at a slower rate compared to light tasks. Specifically, this waves showed an increase from baseline to 125 milliseconds post warning in light activity task, whereas this increase extended from baseline to 0 to 250 milliseconds post warning in moderate activity task. This finding suggests that participants required more time for focused attention, problem-solving, and deep concentration during moderate tasks. Additionally, since these two tasks are different, this discrepancy could be due to the concept that attention involves the allocation of cognitive resources to specific stimuli or tasks.


In our study, we observed a decrease in alpha wave activity, as illustrated in Figure \ref{fig:compare_warnings}, which corresponds to increased attentiveness. Alpha suppression is generally linked to enhanced attention and visual processing. However, our findings reveal no consistent trend in alpha activity between light and moderate tasks after the warnings were issued. Notably, across all three warning types, alpha wave activity in moderate tasks begins at a higher baseline and decreases more gradually compared to light tasks. This desynchronization of alpha waves, as previously established, reflects heightened cognitive engagement and attentiveness.


Another type of brain wave to consider is gamma waves. As shown in Table \ref{table:EEG_metrics}, gamma waves are typically associated with working memory, situational awareness, and memory processes. High gamma wave activity is also closely linked to memory processes. Our findings align with these observations. As depicted in Figure \ref{fig:compare_warnings}, gamma wave activity increased following the warnings for both light and moderate tasks. However, the intensity of gamma waves varied across task activities. After the warnings, gamma waves began at a lower level and rose more gradually during moderate tasks compared to light tasks. This difference may be attributed to the higher cognitive load associated with moderate tasks. The increased cognitive load could act as a barrier, reducing participants' capacity for attention, working memory, and memory processes.


Theta waves are commonly associated with drowsiness or idle states in both children and adults, and they are linked to short-term memory, making them a reliable indicator of mental workload. Additionally, theta waves play a pivotal role in attention processing and working memory. In this study, theta waves show a decreasing pattern in post warnings, as shown in \ref{fig:compare_warnings}. Specifically, during moderate activity tasks, theta waves started at a higher level and decreased at a slower rate compared to light activity tasks. This difference could be due to the improvement in alertness and reduction in drowsiness after warnings. However, the higher cognitive load and complexity of moderate tasks compared to light tasks likely result in a slower reduction of theta waves, indicating that participants require more effort to maintain focus and avoid drowsiness in more demanding tasks (Moderate Activity tasks).

\section{Discussion}
\label{sec5}



\subsection{Relationship Between Reaction Time and EEG-based Cognitive Metrics in High and Low Reaction Performers}

Reaction time to warnings is one of the key indicator of situational awareness, reflecting how quickly participants process and respond to stimuli. However, there is a gap in research exploring the relationship between EEG-derived cognitive metrics and reaction times. In a connected experiment conducted by the research team \citep{sabeti2024augmented}, reaction times were also recorded for the same participants of this study as they responded to warnings in a VR environment. We found that Participant 6 (P06), a 24-year-old male with 4 years of experience in construction, displayed faster reaction times, potentially indicating higher cognitive efficiency.
In contrast, Participant 9 (P09), a 24-year-old male with 5 years of experience in construction, who exhibited consistently slower reaction times, may have struggled with maintaining situational awareness and attention across varying task conditions.
In the following, we will explore the brain wave frequencies of the two participants, hereafter referred to as the high performer (HP: P06) and the low performer (LP: P09), to investigate potential links between their neural activity and reaction times.

In our experimental protocol, participants were subjected to three identical warnings, as detailed in the method section. Analysis of the brain wave data revealed that participants appeared to have adapted to the experimental environment by the second warning, evidenced by the lack of significant difference in post-warning brain wave power patterns between the second and third warnings. This observation aligns with established research on motor and cognitive tasks \citep{dos2017level,schmitz2022enhanced}
, which demonstrates that performance typically stabilizes following an initial adaptation period. Given these findings, our subsequent discussion will focus on the second warning, comparing the  post-warning brain activities of the two high and low performer participants. To facilitate the comparisons, we have visualized the brain activities using heat maps for both participants during both light and moderate tasks for 500 ms post-warnings. These heat maps provide a  spatial representation of neural activity across different brain regions.

\subsubsection{$\alpha$ Band Activity}
\begin{figure}[H]
    \centering
    
    \begin{subfigure}{0.42\textwidth}
        \centering
        \includegraphics[width=\textwidth]{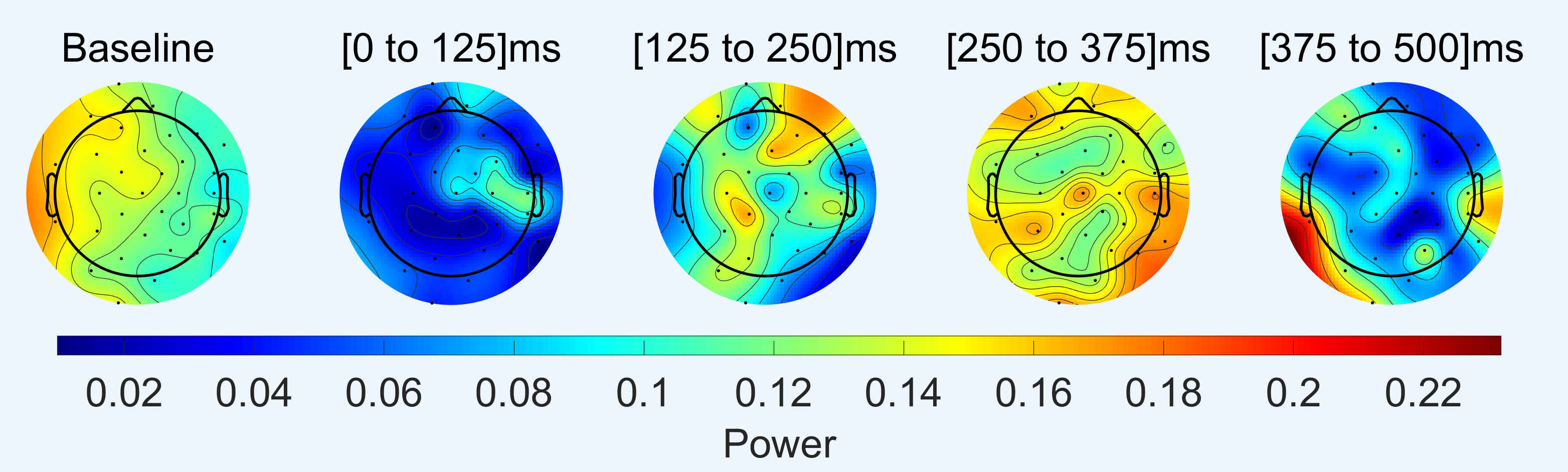}
        \caption{HP Light Activity 
        }
        \label{subfig:P06_alpha_L}
    \end{subfigure}
    \hspace{0.02\textwidth}
    \begin{subfigure}{0.42\textwidth}
        \centering
        \includegraphics[width=\textwidth]{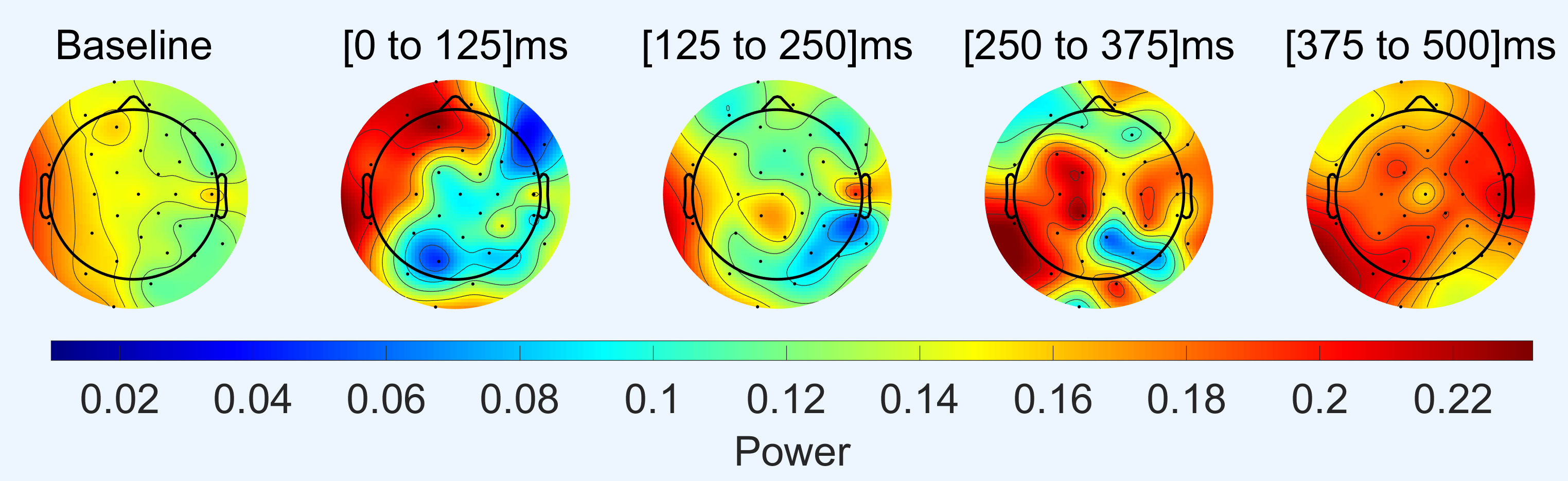}
        \caption{HP Moderate Activity 
        }
        \label{subfig:P06_alpha_M}
    \end{subfigure}
    
    \vspace{0.20cm}
    
    \begin{subfigure}{0.42\textwidth}
        \centering
        \includegraphics[width=\textwidth]{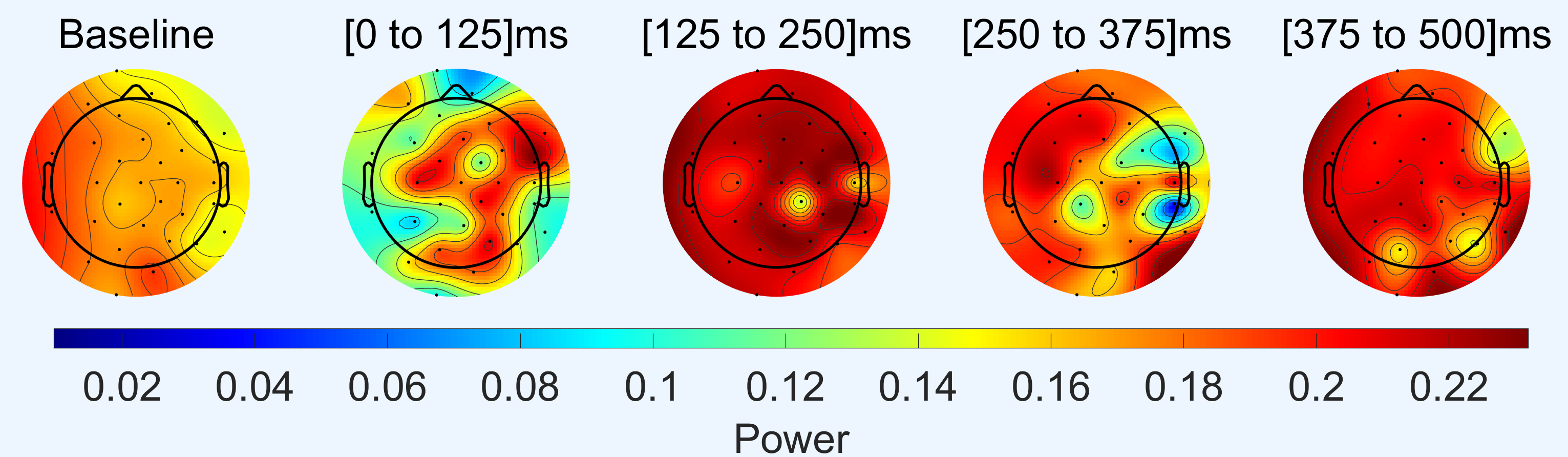}
        \caption{LP Light Activity 
        }
        \label{subfig:P09_alpha_L}
    \end{subfigure}
    \hspace{0.02\textwidth}
    \begin{subfigure}{0.42\textwidth}
        \centering
        \includegraphics[width=\textwidth]{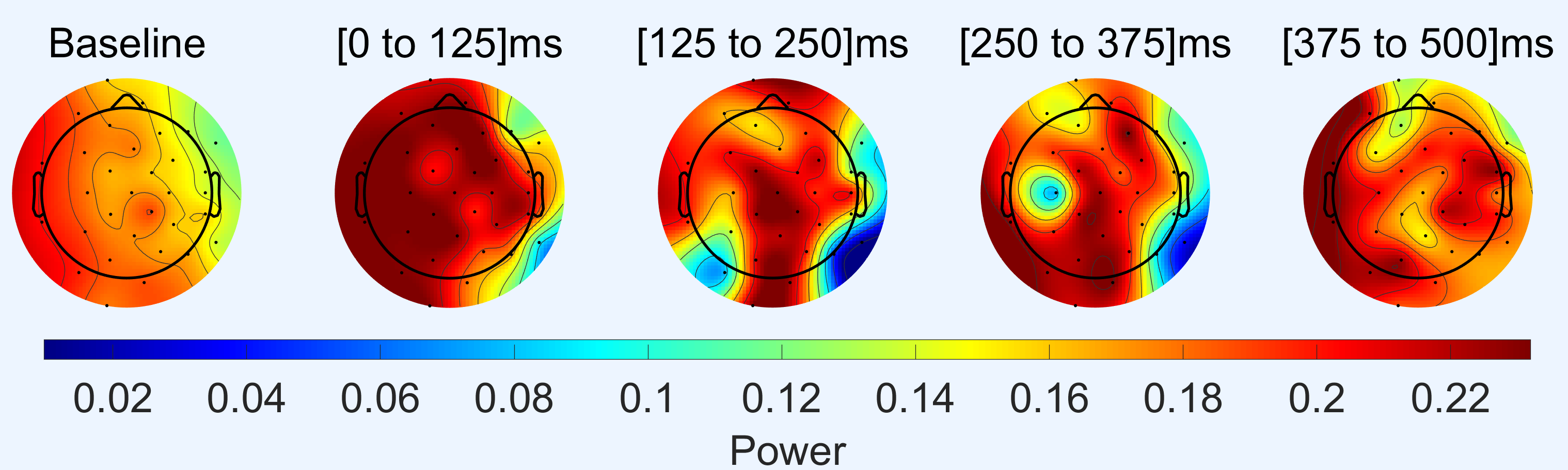}
        \caption{LP Moderate Activity 
        }
        \label{subfig:P09_alpha_M}
    \end{subfigure}
    
    \caption{$\alpha$ band activity of 
    HP (P06) and LP (P09) during Light and Moderate Activity}
    \label{fig:P06&9_alpha_W2_L_M}
\end{figure}

Figure \ref{fig:P06&9_alpha_W2_L_M} illustrates the alpha band activity of the high performer (HP: P06) and the low performer (LP: P09) participants during light and moderate tasks, measured at different time intervals following the warning. The figure highlights the distribution and power of alpha waves across various brain regions, where lower alpha power corresponds to higher attentiveness, and higher alpha power reflects reduced cognitive engagement. 
For HP, as shown in Figure \ref{fig:P06&9_alpha_W2_L_M} (a), the baseline alpha activity is relatively low, suggesting a more attentive state prior to receiving the warning. After the warning delivery, in the 0 to 125 ms post-warning time window, alpha power decreases, indicating increased attentiveness immediately after the warning. However, in the subsequent time windows (125 to 250 ms and beyond), alpha activity increases, suggesting that attentiveness returns to approximately pre-warning baseline status. 
In contrast, as shown Figure \ref{fig:P06&9_alpha_W2_L_M}(c), LP shows consistently higher alpha power both at baseline and throughout the post-warning intervals, suggesting lower attentiveness from the start. In the early time window of 0 to 125 ms post-warning, alpha activity slightly decreases but remains relatively high, particularly in the right temporal and occipital regions, reflecting a slower or weaker response to the warning. As time progresses, during 125 to 250 ms, alpha power increases further and becomes more widespread, extending across the central and parietal regions, indicating that LP's attentiveness continues to decline. LP's pattern of alpha power chanegs remains similar for the moderate activity with increase of alpha power becoming more pornounced and encompassing the rnitire region at at 0-125 ms time window, as shown in Figure \ref{fig:P06&9_alpha_W2_L_M}(d). 


These observations are connected to the reaction times of LP and HP, as LP's higher alpha power and reduced attentiveness likely contribute to slower reaction times, while HP’s greater focus corresponds to faster, more responsive reactions to the warning.
Additionally, the effect of task intensity is evident in the faster onset of increased alpha power in the moderate activity task compared to the light activity task.


\subsubsection{$\theta$ Band Activity}
\begin{figure}[H]
    \centering
    
    \begin{subfigure}{0.42\textwidth}
        \centering
        \includegraphics[width=\textwidth]{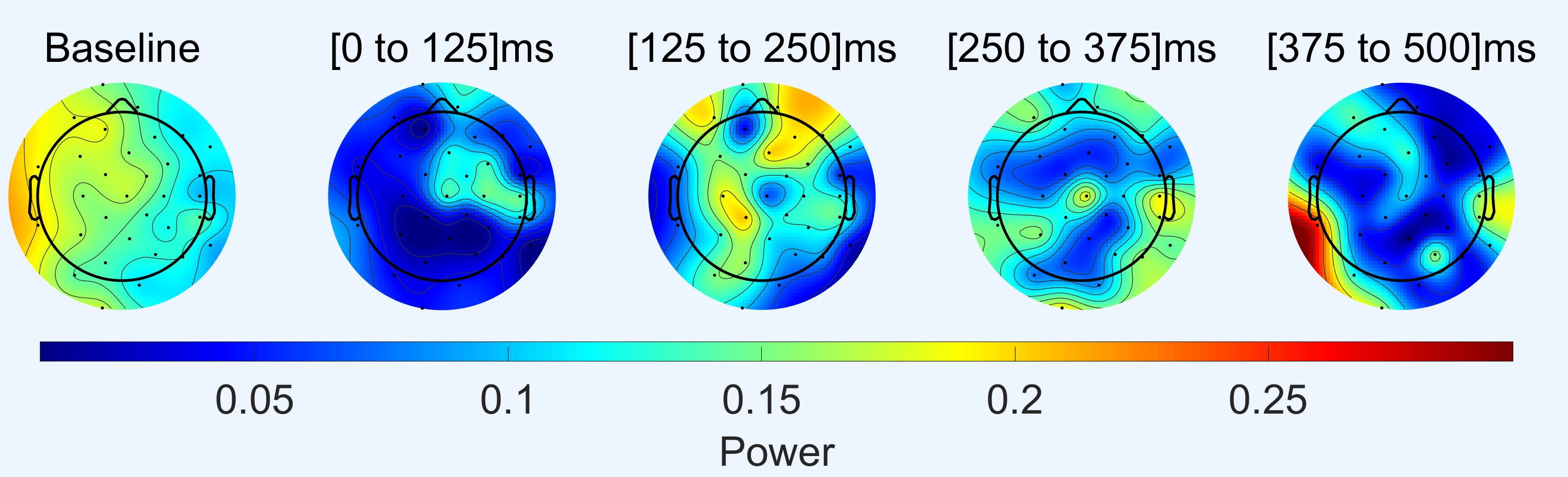}
        \caption{HP Light Activity 
        }
        \label{subfig:P06_theta_L}
    \end{subfigure}
    \hspace{0.02\textwidth}
    \begin{subfigure}{0.42\textwidth}
        \centering
        \includegraphics[width=\textwidth]{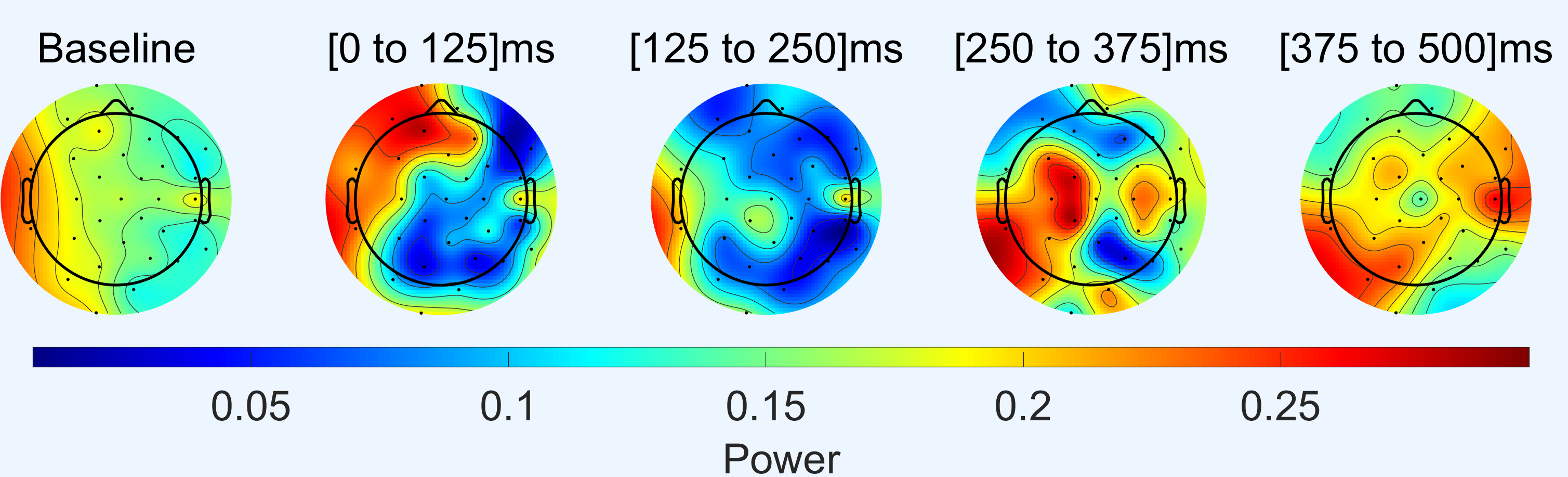}
        \caption{HP Moderate Activity 
        }
        \label{subfig:P06_theta_M}
    \end{subfigure}
    
    \vspace{0.20cm}
    
    \begin{subfigure}{0.42\textwidth}
        \centering
        \includegraphics[width=\textwidth]{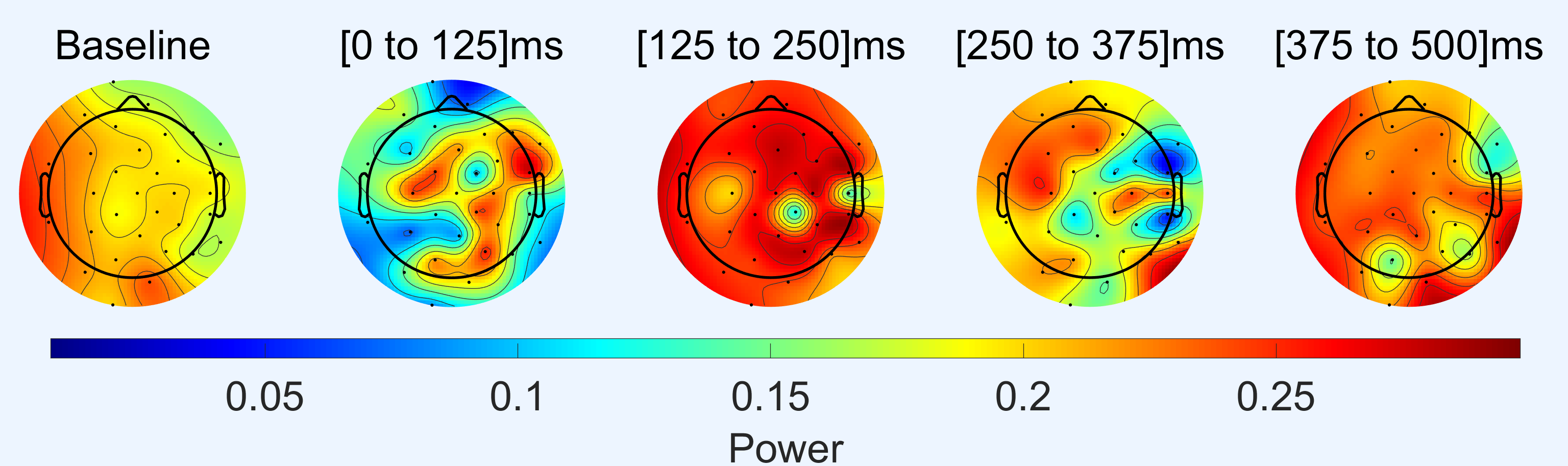}
        \caption{LP Light Activity 
        }
        \label{subfig:P09_theta_L}
    \end{subfigure}
    \hspace{0.02\textwidth}
    \begin{subfigure}{0.42\textwidth}
        \centering
        \includegraphics[width=\textwidth]{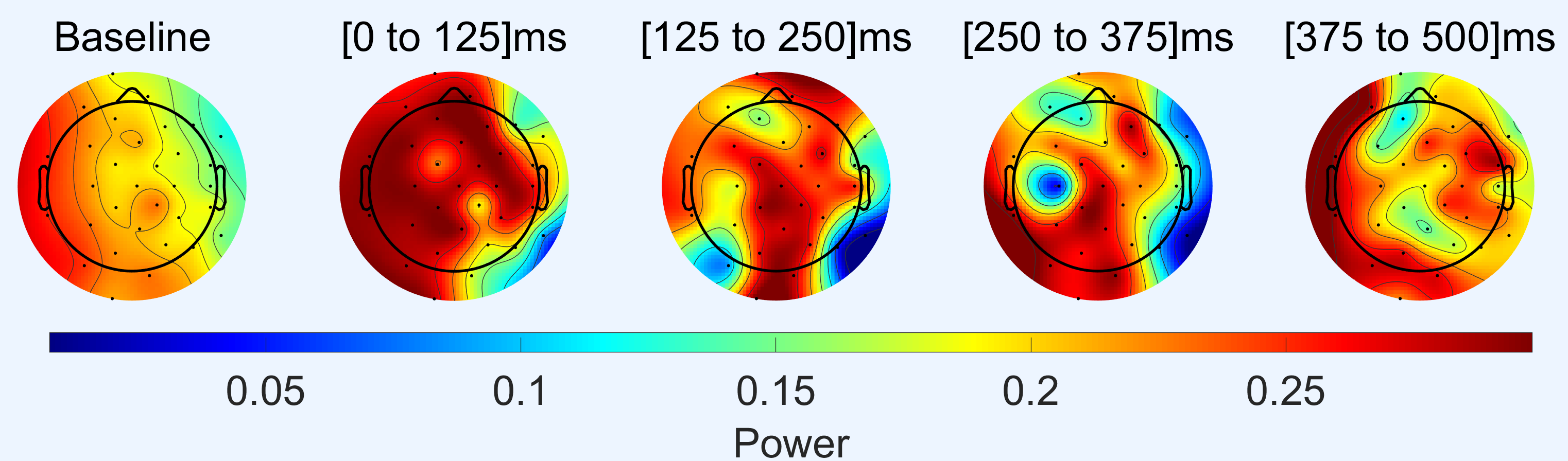}
        \caption{LP Moderate Activity 
        }
        \label{subfig:P09_theta_M}
    \end{subfigure}
    
    \caption{$\theta$ band activity of HP (P06) and LP (P09) during Light and Moderate Activity}
    \label{fig:P06&9_theta_W2_L_M}
\end{figure}

In comparing HP and LP's theta activity following the warning, as shown in Figure \ref{fig:P06&9_theta_W2_L_M}, HP shows a reduction in theta activity during 125 ms post-warning performing the light activity task. However, during the moderate task, HP reallocates cognitive resources, with increased theta activity in the left temporal and frontal lobes—regions associated with working memory and cognitive control. At the same time, there is a decrease in theta activity in the right occipital and parietal regions. This suggests that HP is concentrating cognitive resources on areas critical for task engagement while reducing activity in less task-relevant regions, demonstrating a focused response to the increased task demands. In contrast, LP exhibits a different pattern. During 125 ms post-warning performing the light activity task, there is an increase in theta activity primarily in the right temporal, central, parietal, and occipital regions. For the same time period, with increased task intensity in the moderate activity, this theta activity becomes more widespread, with nearly the entire brain showing elevated levels. This broader activation reflects LP’s heightened effort to stay cognitively engaged, but the widespread increase may suggest a less efficient reallocation of cognitive resources compared to HP. It appears that LP's brain activate more uniformly across regions rather than focusing on specific areas relevant to task performance.

These differences in theta patterns align with the reaction times of both participants, with HP's more efficient theta activation leading to faster responses, whereas LP's less focused activity corresponds to slower reactions.  In addition, the changes in theta activity for both participants corroborate with their alpha activity patterns. This underscores the roles of both alpha and theta waves in cognitive processes related to task complexity and engagement, aligning with the findings of  Klimesch et al. \citep{KLIMESCH1999169}.

\subsubsection{$\beta$ Band Activity}
\begin{figure}[H]
    \centering
    
    \begin{subfigure}{0.42\textwidth}
        \centering
        \includegraphics[width=\textwidth]{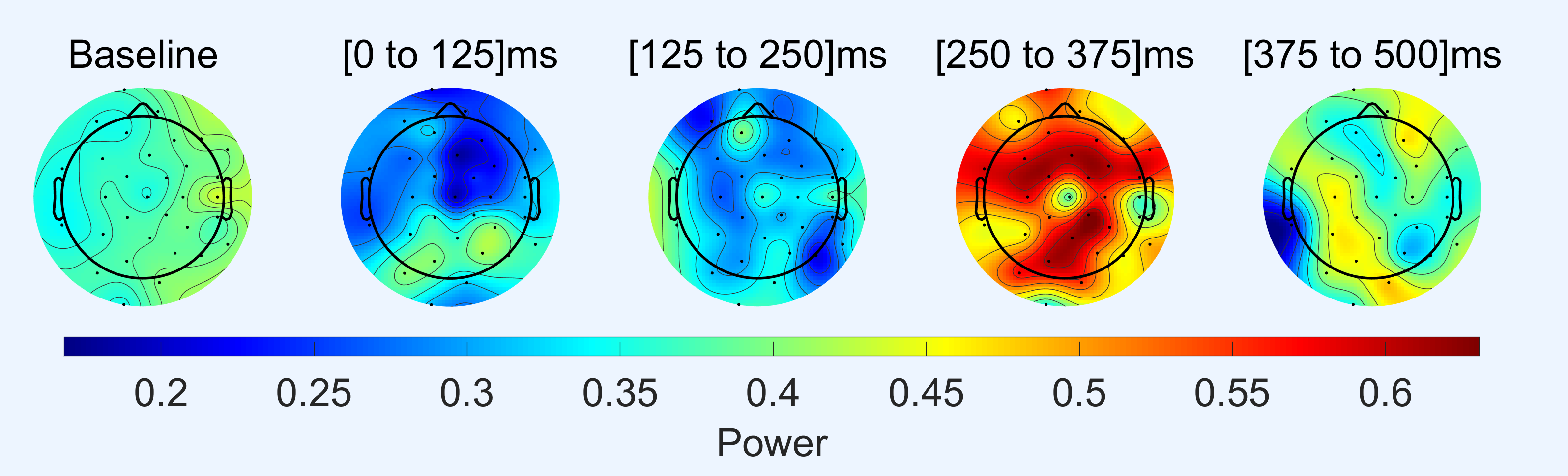}
        \caption{HP Light Activity 
        }
        \label{subfig:P06_beta_L}
    \end{subfigure}
    \hspace{0.02\textwidth}
    \begin{subfigure}{0.42\textwidth}
        \centering
        \includegraphics[width=\textwidth]{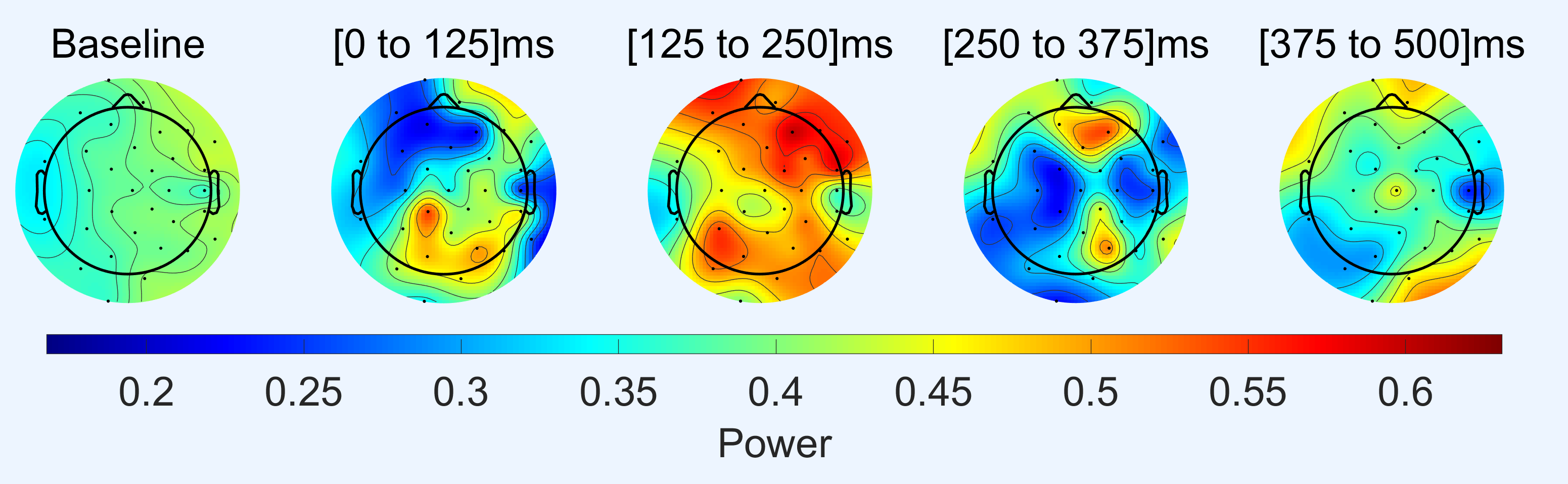}
        \caption{HP Moderate Activity 
        }
        \label{subfig:P06_beta_M}
    \end{subfigure}
    
    \vspace{0.20cm}
    
    \begin{subfigure}{0.42\textwidth}
        \centering
        \includegraphics[width=\textwidth]{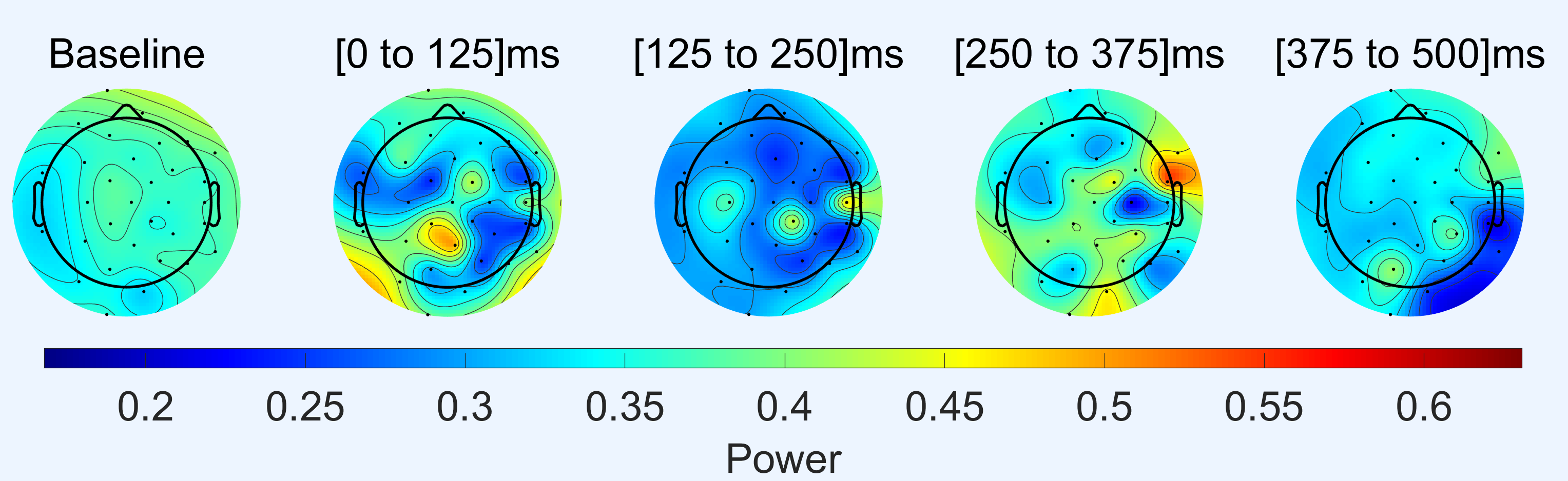}
        \caption{LP Light Activity 
        }
        \label{subfig:P09_beta_L}
    \end{subfigure}
    \hspace{0.02\textwidth}
    \begin{subfigure}{0.42\textwidth}
        \centering
        \includegraphics[width=\textwidth]{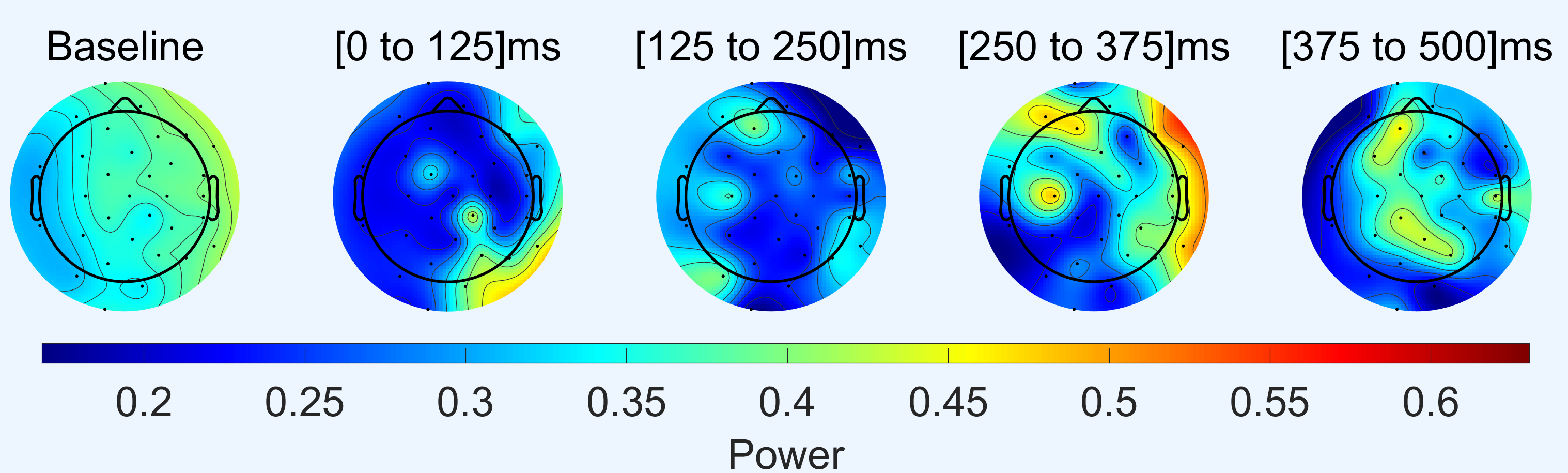}
        \caption{LP Moderate Activity 
        }
        \label{subfig:P09_beta_M}
    \end{subfigure}
    
    \caption{$\beta$ band activity of HP (P06) and LP (P09) during Light and Moderate Activity}
    \label{fig:P06&9_beta_W2_L_M}
\end{figure}

In comparing HP and LP’s beta activity following the warning, as shown in Figure \ref{fig:P06&9_beta_W2_L_M}, HP shows an increased in beta activity during the interval 250 to 375 ms post-warning. However, during the moderate task, HP reallocates cognitive resources, with increased beta activity in the left temporal and frontal lobes during the interval 125 to 250 ms. This pattern indicates that HP's brain reallocates resources to manage the increased task demands, focusing more on regions associated with situational awareness and cognitive control. The increased in beta activity for HP may reflect a strategy of enhancing cognitive focus and reducing distractions to effectively handle the more complex tasks. In contrast, as shown in Figure \ref{fig:P06&9_beta_W2_L_M} (c) and (d), LP exhibits a more pronounced reduction in beta activity during moderate tasks compared to light tasks. In the light activity task, 125 ms post-warning, there is a decrease in beta activity particularly in the temporal, central, and parietal regions, with the most redaction occurring in the interval 125 to 250 ms. During the moderate activity, the entire brain shows decreased beta activity, suggesting that the reduction in beta activity is associated with decrease situational awareness and cognitive efficiency in response to the complexity of the task. Since increase beta wave activity is associated with attention, situational awareness, problem-solving, and deep concentration, this pattern indicates that LP has a lower level of attentiveness and SA compared to HP.

These differences in beta patterns align with the reaction times of both participants, with
HP’s more efficient beta activation leading to faster responses, whereas LP’s reduced attentiveness and situational awareness
results in slower reaction time.  



\subsubsection{$\gamma$ Band Activity}

\begin{figure}[H]
    \centering
    
    \begin{subfigure}{0.42\textwidth}
        \centering
        \includegraphics[width=\textwidth]{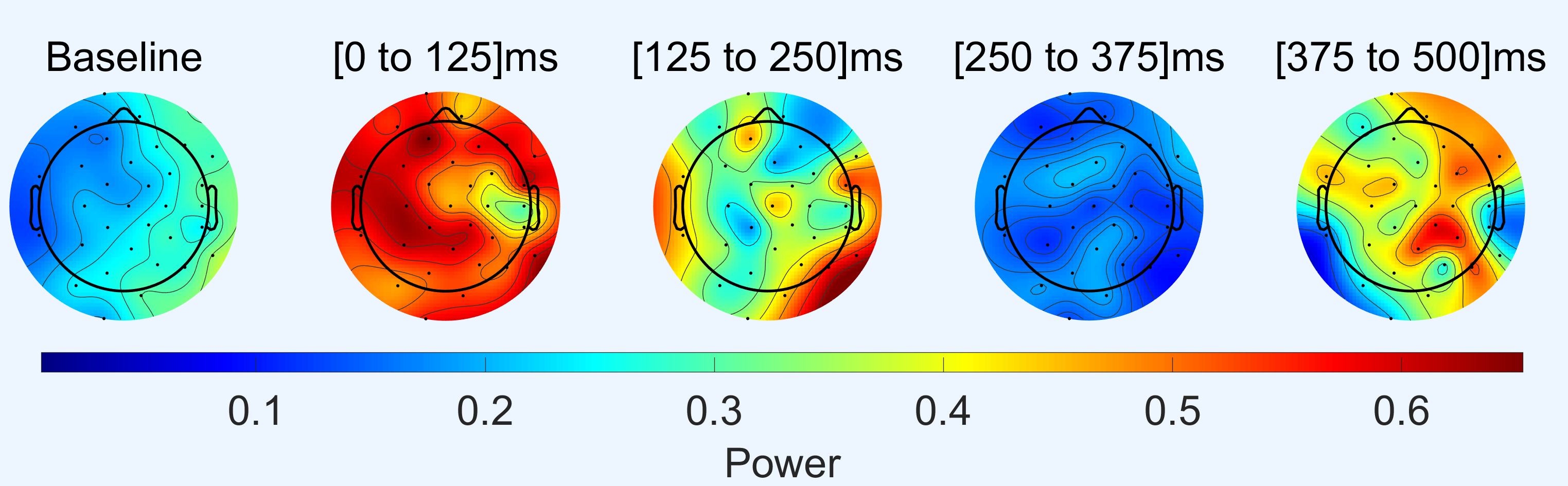}
        \caption{HP Light Activity 
        }
        \label{subfig:P06_gamma_L}
    \end{subfigure}
    \hspace{0.02\textwidth}
    \begin{subfigure}{0.42\textwidth}
        \centering
        \includegraphics[width=\textwidth]{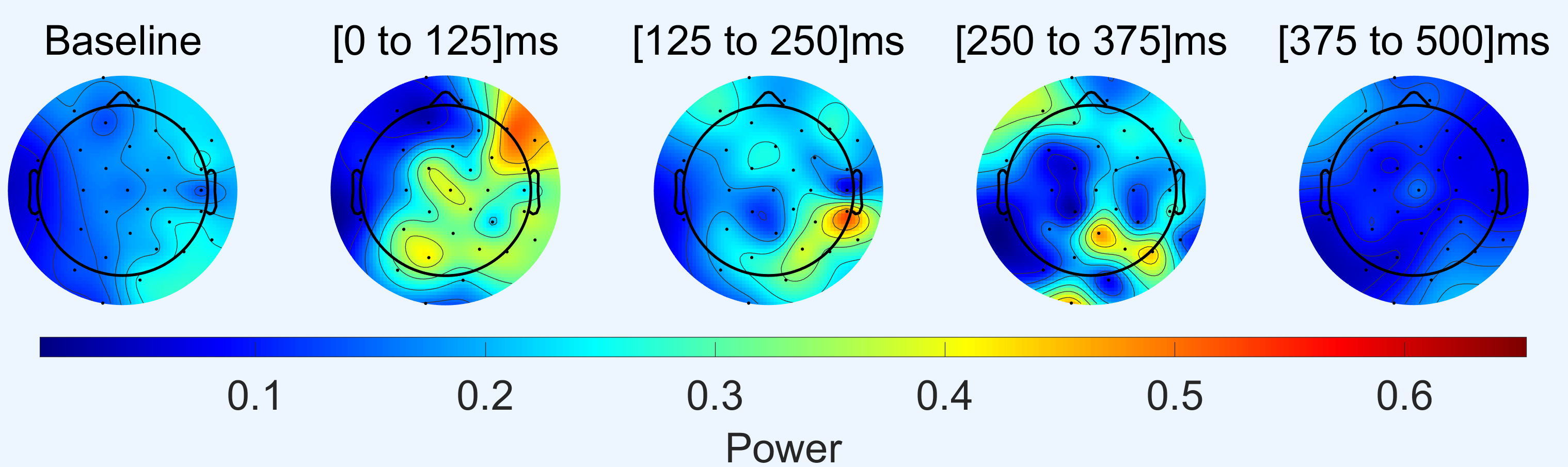}
        \caption{HP Moderate Activity 
        }
        \label{subfig:P06_gamma_M}
    \end{subfigure}
    
    \vspace{0.20cm}
    
    \begin{subfigure}{0.42\textwidth}
        \centering
        \includegraphics[width=\textwidth]{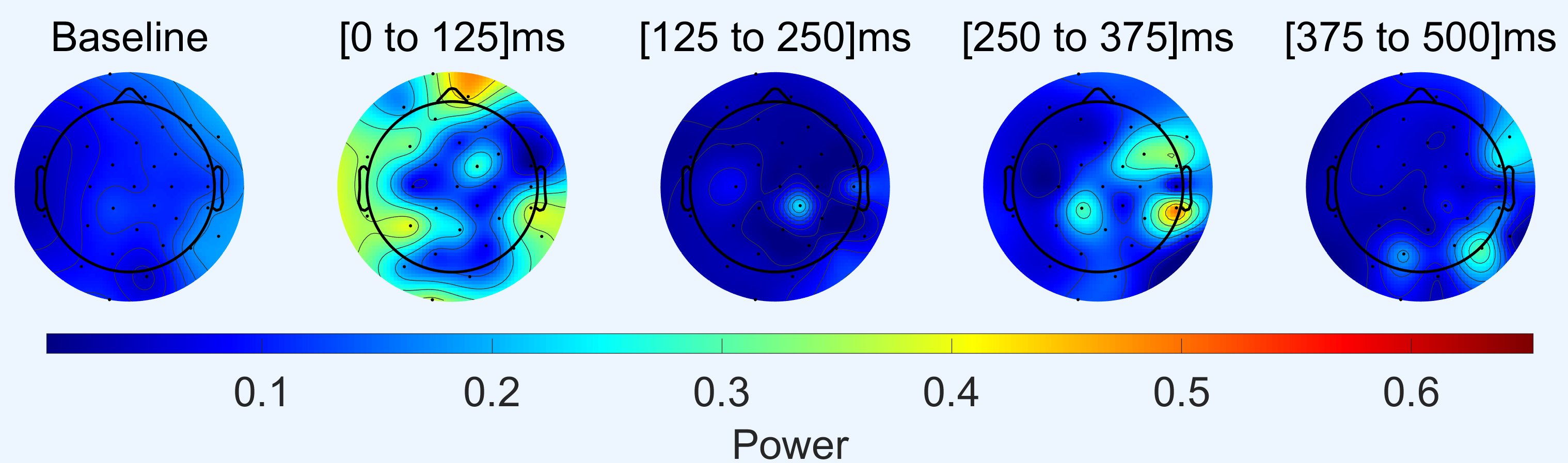}
        \caption{LP Light Activity 
        }
        \label{subfig:P09_gamma_L}
    \end{subfigure}
    \hspace{0.02\textwidth}
    \begin{subfigure}{0.42\textwidth}
        \centering
        \includegraphics[width=\textwidth]{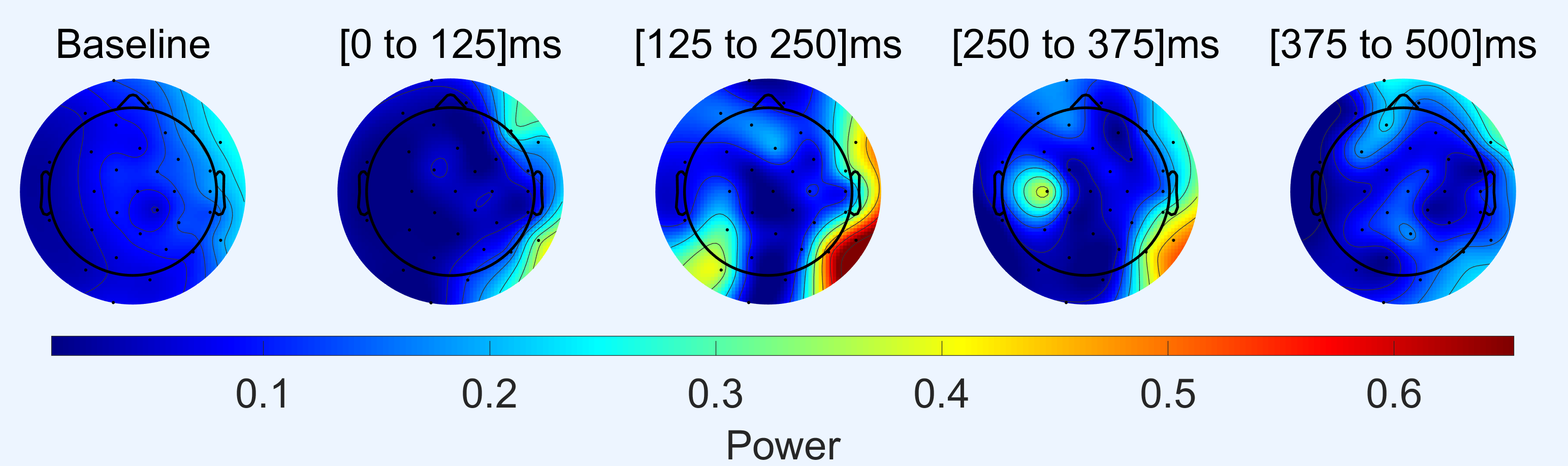}
        \caption{LP Moderate Activity 
        }
        \label{subfig:P09_gamma_M}
    \end{subfigure}
\caption{$\gamma$ band activity of HP (P06) and LP (P09)  during Light and Medium Activity}
    \label{fig:P09&6_gamma_W2_L_M}
\end{figure}

In comparing HP and LP’s gamma activity following the warning, as shown in \ref{fig:P09&6_gamma_W2_L_M}, HP a significant increase in gamma activity during 125 ms post-warning of the light activity task. However, during the medium activity, as shown in Figure \ref{fig:P09&6_gamma_W2_L_M} (b), there is a less pronounced increased in gamma activity, particularly on the right side. This pattern indicates that HP's brain reallocates resources to cope with the increased task demands, focusing on regions associated with attention and cognitive control. The reduction in gamma activity during medium tasks may also reflect a decrease in situational awareness as the task complexity increases. In contrast, as shown in Figure \ref{fig:P09&6_gamma_W2_L_M} (c) and (d), LP  exhibits increase in gamma activity sooner during light tasks compared to medium tasks. In the light activity task, during 125 ms post-warning, there is an improvement in gamma activity particularly in the temporal and partial frontal regions. However, during the medium activity, the entire brain shows decreased gamma activity, suggesting that the reduction in gamma activity is associated with the cognitive load and increased task complexity. However, between 125 ms and 250 ms post-warning there is a slight improvement in the right side of the brain.

These differences in gamma patterns align with the reaction times of both participants, with
HP’s more efficient gamma activation leading to faster responses, whereas LP’s increased cognitive load and reduced situational awareness
results in slower reaction time. 

\subsection{Neuro-physiological Sensing for Enhanced Work Zone Safety}

Our study's use of EEG technology to assess neurophysiological responses to AR-based safety warnings represents a significant advancement in work zone safety research. Traditional safety measures often focus on physical hazards, neglecting the cognitive states of workers that are crucial for maintaining situational awareness and preventing accidents. Our findings demonstrate the feasibility and value of using EEG to objectively measure these cognitive states.
The EEG data collected during our simulations revealed significant changes in brain activity patterns in response to AR multi-sensory warnings. Specifically, we observed increased activity in regions associated with attention and situational awareness when participants were exposed to these warnings. This aligns with previous research by Ramos-Hurtado et al. \cite{ramos2022proposal} and Lukosch et al. \cite{lukosch2015providing}, who found that AR can improve workers' understanding of safety risks. However, our study extends these findings by providing objective, neurophysiological evidence of enhanced cognitive processes.
Importantly, our results also highlighted the impact of workload on the effectiveness of AR warnings. Under high physical workload conditions, the EEG data showed a diminished response to AR warnings, suggesting that cognitive resources may be diverted from processing these alerts when workers are physically strained. This finding supports the work of Kim et al. \cite{kim2022construction} and Zhang et al. \cite{zhang2023digital}, who noted the influence of physical workload on situational awareness. Our study, however, provides more nuanced insights into this relationship in the context of AR-assisted work zones.

The discrepancies we observed between subjective assessments of situational awareness and objective EEG-based metrics underscore the importance of using neurophysiological measures in safety research. While participants generally reported improved situational awareness with AR warnings, the EEG data revealed more variability, particularly under high workload conditions. This highlights a potential limitation of relying solely on self-reported measures and emphasizes the value of objective neurophysiological monitoring.


Furthermore, our research opens up new possibilities for continuous monitoring of workers' cognitive states in real-world work zones. While our study was conducted in a simulated environment, the methodology we developed could be adapted for field use. This could involve the development of wearable EEG devices integrated into safety equipment, such as hard hats. Such a system could provide real-time alerts to workers and supervisors when cognitive states indicative of reduced situational awareness or increased accident risk are detected.

Moreover, the use of such advanced safety technology could have insurance implications. By demonstrating a commitment to proactive safety measures and reducing the risk of accidents, construction companies could potentially negotiate lower insurance premiums. The continuous data collection and analysis could provide tangible evidence of improved safety practices and reduced incident rates, making a compelling case for insurance providers to offer more favorable terms. This not only benefits the companies financially but also underscores the value of investing in innovative safety solutions to protect workers.

\subsection{Implications for Workforce Training and Task Matching}

The observed variability in EEG responses to AR warnings under different workload conditions suggests that workers may benefit from personalized training programs. By utilizing a combination of AR and EEG technologies in a Virtual Reality training environment, more effective and targeted training scenarios could be created. These scenarios could be designed to replicate the cognitive demands of actual work zones, allowing workers to develop and refine their situational awareness skills in a safe, controlled setting.  This approach aligns with the findings of Paes et al. \cite{paes2024optical}, who demonstrated the importance of using AR technology for safety training.

Our EEG data revealed that some participants maintained high levels of situational awareness even under high workload conditions, while others showed significant decreases. This variability indicates that individual differences play an important role in how workers respond to AR safety systems. By incorporating EEG monitoring into VR training sessions, it would be possible to identify workers who may need additional support or training, including indivduals with disabilities, to maintain optimal situational awareness in challenging conditions. 

Furthermore, the EEG data collected during our study provides insights into the cognitive load experienced by workers when processing AR warnings. This information could be valuable for task matching within work zones. For instance, workers who demonstrate a high capacity for processing multiple sensory inputs while maintaining low cognitive load might be better suited for roles that require constant interaction with AR safety systems. Conversely, those who show increased cognitive load with AR warnings might be more effective in roles with less reliance on these systems.

Additionally, this approach holds promise for workers with disabilities. By quantifying their cognitive measures through continuous EEG monitoring, it is possible to identify the most suitable tasks for these individuals. Tailoring task assignments based on cognitive strengths allows workers with disabilities to become more active and integrated within the workforce. This not only enhances their productivity and job satisfaction but also fosters an inclusive work environment. 


\section{Conclusion}
\label{sec6}

This study aimed to assess situational awareness, attention, and cognitive workload experienced by construction workers receiving safety warnings in roadway work zones. By leveraging EEG sensing technology within a high-fidelity virtual reality environment, we have gained valuable insights into the neurological responses of workers to multi-sensory warnings during light and moderate-intensity tasks. Our findings reveal distinct patterns in brain activity across different task intensities and in response to warnings. Situational awareness indicators, including beta and gamma waves, demonstrated increases post-warning, with varying intensities between task types. Furthermore, attention indicators such as alpha and theta waves showed a consistent decrease following warnings, with moderate-intensity tasks exhibiting higher initial levels and slower rates of decrease. Analyzing the correlation between brain activity bands (alpha, beta, and gamma) and NASA-TLX factors, we found no consistent pattern in the direction of correlation across different warnings and tasks.

The human-sensing model developed in this study represents a significant advancement in the field. By incorporating brain frequency band characteristics, it offers a more nuanced understanding of workers' cognitive states during AR-enabled warning reception. This model not only enhances our ability to assess situational awareness, attention, and cognitive workload but also clarifies the relationship between these factors and physical effort.  This research contributes to the growing body of knowledge on construction worker safety, particularly in high-risk environments such as roadway work zones. By deepening our understanding of workers' cognitive responses to warnings and task intensities, we pave the way for more effective, personalized safety measures. As we continue to advance in this field, the ultimate goal remains clear: to significantly reduce accidents and fatalities in construction work zones through evidence-based, neurologically-informed safety practices and technology development.

We acknowledge a few limitations of this study that should be addressed in future research. The study was conducted in a simulated virtual reality environment, which, while allowing for controlled low-risk experimentation, may not fully capture the complexity and unpredictability of real-world work zones.  Furthermore, our study focused primarily on immediate cognitive responses to AR warnings, and did not assess the long-term effects of using such safety systems. Future research should consider longitudinal studies to evaluate how workers' responses and performance may change over extended periods of AR system use. Moreover, while we examined different workload conditions, our experimental design may not have captured the full range of physical and cognitive demands experienced by workers. Finally, our study was limited in its exploration of individual differences. Factors such as age, prior experience with AR technologies, disabilities, or cognitive styles could influence how workers respond to AR warnings. Future studies should aim to investigate these individual differences to provide an understanding of AR effectiveness across diverse worker populations.

\section{CRediT Authorship Contribution Statement}

\textbf{Fatemeh Banani Ardecani:} Conceptualization, Methodology, Data curation, Formal analysis, Writing – original draft, Writing – review \& editing, Visualization. 
\textbf{Amit Kumar:} Data curation, Writing – review \& editing. 
\textbf{Sepehr Sabeti:} Data curation, Writing – review \& editing. 
\textbf{Omidreza Shoghli:} Conceptualization, Methodology,  Visualization,  Writing – review \& editing, Supervision, Funding acquisition.

\section{Acknowledgments}
This research was supported in part by the National Science Foundation under Award Number 1932524. We express our sincere gratitude to all study participants for their time and effort.

\bibliographystyle{elsarticle-num-names} 
\bibliography{Refs}

\end{document}